%% file: main.tex
\documentclass[lettersize,journal]{IEEEtran}
\usepackage{amsmath,amsfonts}
\usepackage{algorithmic}
\usepackage{array}
\usepackage[caption=false,font=normalsize,labelfont=sf,textfont=sf]{subfig}
\usepackage{textcomp}
\usepackage{stfloats}
\usepackage{url}
\usepackage{verbatim}
\usepackage{graphicx}
\usepackage{mathrsfs}
\usepackage{color,xcolor}
\usepackage{multirow} 

\usepackage{amsthm}
\usepackage{amssymb}

\usepackage[linesnumbered,lined,ruled]{algorithm2e}
\usepackage{arydshln} 
\usepackage{paralist}
\usepackage{pifont}
\def\BibTeX{{\rm B\kern-.05em{\sc i\kern-.025em b}\kern-.08em
    T\kern-.1667em\lower.7ex\hbox{E}\kern-.125emX}}
\usepackage{balance}
\usepackage{hyperref}
\usepackage[compress]{cite}

\usepackage{tcolorbox}

\begin{document}
\title{\huge{FlexFL: Heterogeneous Federated Learning via APoZ-Guided Flexible Pruning in Uncertain Scenarios}}

\author{Zekai~Chen, 
        Chentao~Jia,
        Ming~Hu,~\IEEEmembership{Member,~IEEE,}
        Xiaofei~Xie,~\IEEEmembership{Member,~IEEE,}
        Anran~Li,~\IEEEmembership{Member,~IEEE,}
        and~Mingsong~Chen,~\IEEEmembership{Senior Member,~IEEE}
\thanks{
The authors Zekai~Chen,  Chentao~Jia, and Mingsong Chen are with the MoE Engineering Research Center of Hardware/Software Co-design Technology and Application at East China Normal University, Shanghai, 200062, China.
The authors Ming Hu and Xiaofei Xie are with Singapore Management University, Singapore.
The author Anran Li is with Yale University, USA.
Ming Hu (hu.ming.work@gmail.com) is the corresponding author.
}
}


\maketitle

\input{abstract}
\input{introduction}
\input{background}
\input{approach}

\input{experiment}

\input{conclusion}

\input{ack}


\bibliographystyle{IEEEtran}

\bibliography{references}

\end{document}

%% file: abstract.tex
\begin{abstract}

Along with the increasing popularity of Deep Learning (DL) techniques, more and more Artificial Intelligence of Things (AIoT) systems are adopting federated learning (FL) to enable privacy-aware collaborative learning among AIoT devices. 
However, due to the inherent data and device heterogeneity issues, existing FL-based AIoT systems suffer from the model selection problem. 
%
Although various heterogeneous FL methods have been investigated to enable collaborative training among heterogeneous models, there is still a lack of 
i) wise heterogeneous model generation methods for devices, ii) consideration of 
uncertain factors, and iii) performance guarantee for large models, thus strongly limiting the overall FL performance.   
To address the above issues, this paper introduces a novel heterogeneous FL framework named FlexFL. 
By adopting our  Average Percentage of Zeros (APoZ)-guided flexible pruning strategy, FlexFL can effectively derive best-fit models for heterogeneous devices to explore their greatest potential.  Meanwhile, our proposed adaptive local pruning strategy allows AIoT devices to prune their received models 
according to their varying resources within uncertain scenarios. 
Moreover, based on self-knowledge distillation, FlexFL can enhance the inference performance of large models by learning knowledge from small models. 
Comprehensive experimental results show that,  compared to state-of-the-art heterogeneous FL methods, FlexFL can significantly improve the overall inference accuracy by up to 14.24\%.


\end{abstract}


%% file: introduction.tex
\section{Introduction}

Along with the prosperity of Artificial Intelligence  (AI) and the Internet of Things (IoT), 
Federated Learning (FL)~\cite{FedAvg,li2023towards,hu2024fedcross,mo2021two,li2021sample,hu2024fedmut,yan2023have}
is becoming a mainstream
distributed Deep Learning (DL) paradigm in the design 
Artificial Intelligence of Things (AIoT) systems~\cite{zhang2020efficient,hao2019fpga,hu2023aiotml}, since it enables collaborative learning among devices without compromising their data privacy. So far, FL has been widely investigated in various AIoT applications, such as
intelligent transportation \cite{manias2021making,yertss2023},
edge-based mobile computing~\cite{wang2019edge,li2021hermes},  
real-time control~\cite{li2019smartpc,hu2023gitfl}, and healthcare systems~\cite{wu_iccad2021,nguyen2022federated}.
Typically, an FL-based AIoT system is based on a client-server architecture involving a cloud server and numerous AIoT devices. 
In each FL training round, the cloud server first dispatches the latest global model to multiple selected (activated) devices for local training and then aggregates the trained local models to update the global model. Since 
the communication between the cloud server and devices is based on model gradients, FL enables knowledge sharing among AIoT devices without privacy leaks. 


Although existing FL methods are promising for sharing knowledge among devices, they are not well-suited for large-scale AIoT applications involving various heterogeneous devices with different available resources \cite{singh2019collaborative}. 
This is because traditional FL methods assume that all the device models are of the same architecture. According to the Cannikin Law,  only the models best fit for the weakest devices can be used for FL training. Typically, such models are of small sizes with limited inference capability, thus strongly 
suppressing the potential FL learning performance. 
To maximize the knowledge learned on heterogeneous devices, 
various heterogeneous FL methods have been investigated to use heterogeneous models for local training, which can be mainly classified into two categories, i.e., 
%
%
{\it completely heterogeneous methods} and {\it partially heterogeneous methods}.
Specifically, completely heterogeneous methods~\cite{cho2022heterogeneous} apply  
Knowledge Distillation (KD) strategies~\cite{gou2021knowledge,hinton2015distilling,park2019relational}
on heterogeneous models with totally different structures for knowledge sharing,  
%
while partially heterogeneous methods~\cite{heterofl,depthfl,jia2023adaptivefl,ScaleFL} derive heterogeneous models from the same large global model for local training and knowledge aggregation. 
As an example of partially heterogeneous methods, for a given large global model, HeteroFL~\cite{heterofl} can generate heterogeneous models to fit devices by pruning the parameters of
each model layer. 


When dealing with real-world AIoT applications, existing heterogeneous FL methods greatly suffer from the following three problems: 
\ding{182} low-performance heterogeneous models derived by unwise model pruning strategies, 
\ding{183} inefficient or ineffective local training within uncertain scenarios, 
and \ding{184} low inference performance of large models caused by resource-constrained scenarios.
Specifically, existing methods generate
heterogeneous models coarsely by pruning the parameters of each model layer with the same ratio or directly removing the entire layer. 
Without considering the different functions of parameters within layers, such unwise pruning strategies severely limit the performance of heterogeneous models.
Meanwhile, when encountering  various uncertainty factors, such as 
hardware performance fluctuations caused by
process variations ~\cite{hu2020quantitative,hu2023gitfl}, 
dynamic resource utilization (e.g., available memory size),
traditional FL may fail in local training since they assume static device resources during FL training. 
Due to the inaccurate estimation of available device resources, the overall training performance can be deteriorated. 
%
Furthermore, 
within a large-scale AIoT application, typically large models cannot be accommodated by most resource-constrained devices.
As a result, the small amount of training data will inevitably influence the inference capability of large models.
Therefore, \textit{how to wisely generate high-performance heterogeneous models to fit for uncertain scenarios is becoming an urgent issue in heterogeneous FL design.
}

Intuitively, to achieve high-performance pruned models,  a model pruning method should
delete the least significant neurons first. As a promising measure, activation information can be used to evaluate the importance of neurons, where neurons with more activation times have greater importance. In other words, if a model layer consists of more neurons with fewer activation times, it has more parameters to be pruned. 
According to \cite{hu2016network}, the Activation Percentage of Zeros (APoZ)  can be used to measure the percentage of zero neuron activation times under the Rectified Linear Unit (ReLU) mapping. Therefore, the higher the APoZ score of a model layer, the higher the pruning ratio we can apply to the layer. 
%
%
Based on this motivation, 
this paper proposes
a novel heterogeneous FL approach named FlexFL, which 
utilizes the APoZ scores of model layers to perform finer-grained pruning to 
generate high-performance heterogeneous to best fit their target devices for high-quality local training. 
To accommodate various uncertain scenarios, 
FlexFL allows devices to adaptively prune their received models according to their available resources.
Meanwhile, based on a self-KD-based training strategy, FlexFL 
enables large models to learn from small models, thus improving their inference performance. 
Note that in FlexFL, the small models are derived from large models, and the self-KD-based training is only performed by devices. In this way, FlexFL can effectively explore the greatest potential of devices, thus improving the overall FL training performance. 
This paper makes the following four major contributions: 
\begin{itemize}
    \item We propose an APoZ-guided flexible pruning strategy to wisely generate heterogeneous models best for devices.  
    \item We design an adaptive local pruning strategy to enable devices to further prune their local models to adapt to varying 
 available resources within uncertain scenarios.

    \item We present a self-KD-based local training strategy that utilizes the knowledge of small models to enhance the training of large models.

    \item We perform extensive experiments based on 
    simulation and real test-beds to evaluate the performance of FlexFL. 

\end{itemize}


%% file: background.tex
\section{Background And Related Work}\label{section: background}

\subsection{Background}
\subsubsection{Federated Learning}
Federated Learning (FL) is a distributed machine learning approach that addresses data privacy protection and decentralization issues. 
Traditional FL framework usually consists of a server and multiple devices. In FL training, the server maintains a global model and dispatches it to the selected devices for local training in each round. Each device then trains locally on its own data and uploads the trained model to the server after training. Finally, the server aggregates all the received models to generate a new global model. Specifically, the optimization objective of FL is based on FedAvg~\cite{FedAvg}, which is defined as follows:
\begin{equation}
    \begin{split}
        &\min_{w}F(w)=\frac{1}{|D|}\sum_{k=1}^{|D|}f_{k}(w),\\
        &\text {s.t.,} f_{k}(w)=\frac{1}{|{\mathcal{D}_k}|}\sum_{i=1}^{|{\mathcal{D}_k}|}\ell\left(w,\langle x_{i},y_{i}\rangle\right),
        \nonumber
    \end{split}
\end{equation}
where $|D|$ denotes the number of devices, and the function $f_{k}(w)$ is the loss value of the model on device $k$, $|{\mathcal{D}_k}|$ denotes the data set size in device $k$ and $\ell$ denotes the loss function (e.g., cross-entropy loss), and $w$ is the model parameter, as well as optimization objective, $x_i$ and $y_i$, are the samples and the corresponding labels, respectively.


\subsubsection{Model Pruning}
In the field of machine learning and deep learning, model pruning is a technique to reduce model complexity and computational resource requirements by reducing redundant parameters and connections in neural network models. The main objective of model pruning is to achieve a more compact and efficient model without significantly sacrificing its performance. Initially, the trained model is analyzed to identify parameters or connections that contribute less to the overall model performance. These parameters are considered redundant and can be pruned without affecting the model's performance. 
Common approaches include magnitude-based pruning ~\cite{han2015learning}, which removes parameters with small weights; sensitivity-based pruning ~\cite{mozer1988skeletonization}, which measures the impact of each parameter on the model's output; and structured pruning  ~\cite{he2019filter}, which removes entire neurons or channels.

APoZ (Activation Percentage of Zeros)~\cite{hu2016network} is a metric used in model pruning to quantify the sparsity level of neural network activations. It measures the percentage of zero activations in a layer or network after applying a pruning technique. A high APoZ score indicates that a large proportion of activations in the network are zero, indicating that the network has achieved significant sparsity. In essence, APoZ provides a quantitative measure of sparsity, allowing us to assess the impact of pruning methods on neural network architectures and optimize pruning strategies to achieve the desired trade-off between model size and performance.

\begin{figure*}[t]
    \centering		
    \includegraphics[width=0.8\textwidth]{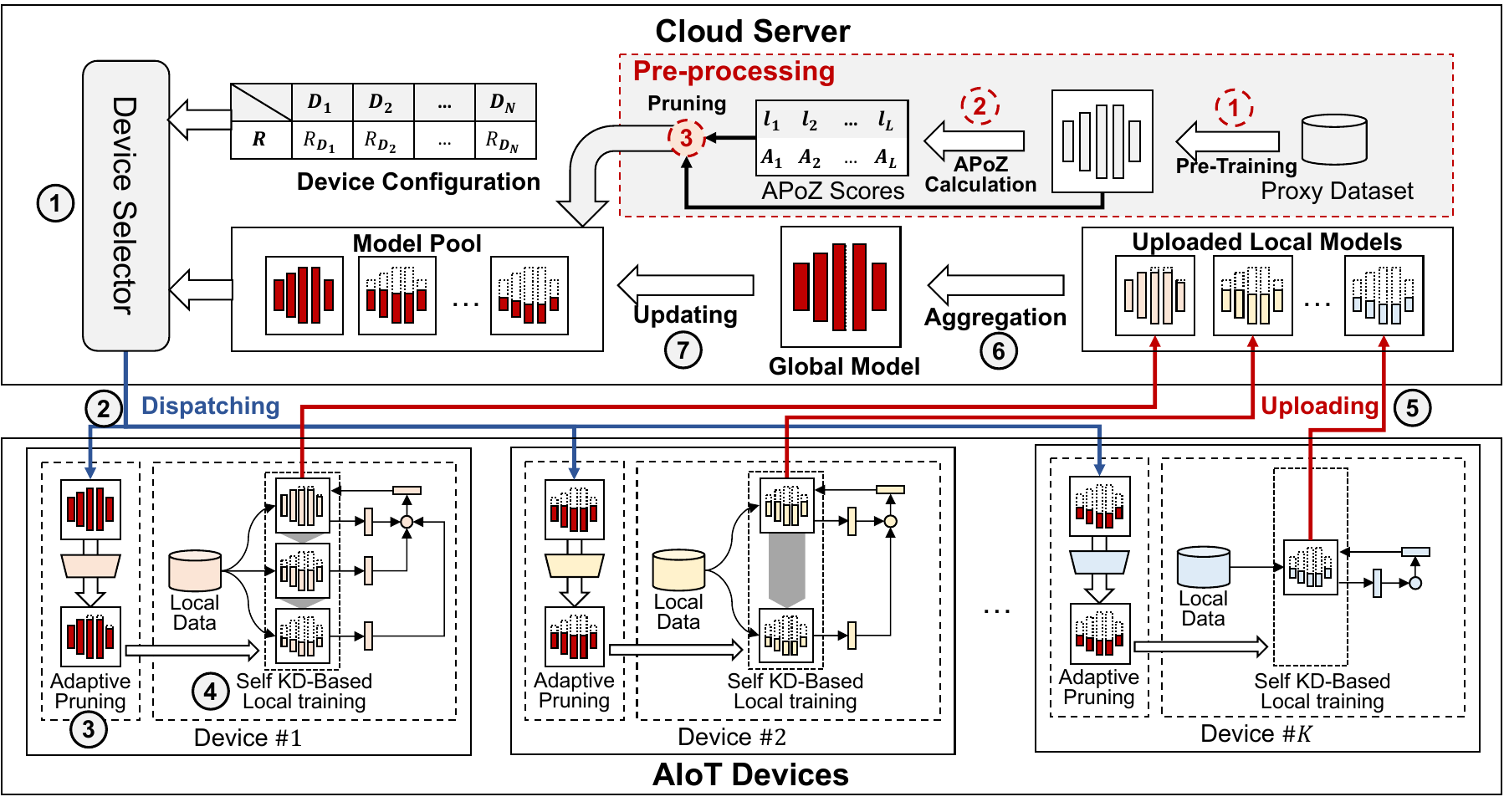}
\vspace{-0.2in}
 \caption{Framework and workflow of FlexFL.}
	\label{fig: framework}
\vspace{-0.2in}
\end{figure*}

\subsection{Model Heterogeneous Federated Learning}
Model heterogeneous FL~\cite{depthfl,heterofl, ScaleFL,liu2023adapterfl} has a natural advantage in solving systemic heterogeneity. Different from traditional FL, model heterogeneous FL usually maintains some models of different sizes on the cloud server, so as to better deal with the heterogeneous resources of different devices.
The current model heterogeneous FL methods can be classified into three categories, i.e., width-wise pruning, depth-wise pruning, and two-dimensional pruning.
For width-wise pruning, in~\cite{heterofl}, Diao \textit{et al.} proposed HeteroFL, which alleviated the problem of device system heterogeneity by tailoring the model width and conducted parameter-averaging over heterogeneous models. Similarly, in~\cite{fjord}, Horvath \textit{et al.} proposed FJoRD, which used the Ordered Dropout mechanism to extract the lower footprint submodels.
For depth-wise pruning, DepthFL~\cite{depthfl} prunes the later parts of deeper networks to reduce the number of network parameters. In InclusiveFL~\cite{inclusivefl}, Liu \textit{et al.} proposed a layer-wise model pruning method with momentum knowledge distillation to better transfer knowledge among submodels.
For two-dimensional pruning, in ScaleFL~\cite{ScaleFL}, Ilhan \textit{et al.} introduced a pruning method that scales from both width and depth dimensions, aiming to balance the proportions of model width and depth. Additionally, it incorporates skip connections to facilitate connections between shallower models and the network's classification layers.
\textit{However, the existing approaches seldom consider the characteristics of each model and the differences in neuronal activation distribution on different datasets, and most of them adopt a fixed pruning method while ignoring the different model architectures. 
Moreover, existing methods largely lack consideration of device-related resource uncertainties in real-world environments. Most of them are studied under the assumption of fixed device resources, which deviates from the dynamic nature of AIoT scenarios.}

To the best of our knowledge, FlexFL is the first attempt to utilize a flexible pruning strategy to generate heterogeneous models in FL under resource uncertainty scenarios.
Using APoZ scores and the number of parameters of each layer, FlexFL can generate higher-performance heterogeneous models for local training.
To deal with various uncertain and resource-constrained scenarios, FlexFL integrates an adaptive local pruning mechanism and self-KD-based local training strategy, which enables devices to adaptively prune their received model according to their available resources and effectively improves the performance of FL training. 

%

%% file: approach.tex
\section{Our FlexFL Approach}\label{section: approach}

\subsection{Overview of FlexFL}
Figure \ref{fig: framework} presents the framework and workflow of FlexFL, which consists of two stages, i.e., the \textit{pre-processing stage} and the \textit{FL training stage}.
FlexFL maintains a large model as the global model and generates multiple heterogeneous models for local training based on the global model.
The pre-processing stage aims to calculate the APoZ scores of each layer, which are used for model pruning to generate heterogeneous models.
The FL training stage aims to train the multiple heterogeneous models, which are pruned from the global model.

As shown in Figure~\ref{fig: framework}, in the pre-processing stage, the cloud server maintains a proxy dataset to pre-train the global model and calculates APoZ scores for each layer according to neuron activation.
Since APoZ calculation does not require the model to be fully trained, the proxy dataset requires only a small amount of data compared to the training dataset.
The cloud server then prunes the global model to generate multiple heterogeneous models based on calculated APoZ scores.
Specifically, the workflow of the pre-processing stage consists of three steps as follows:
\begin{itemize}
    \item \textbf{Step 1: Pre-training.} Since the APoZ scores are calculated based on a trained model, the cloud server uses a proxy dataset to train the global model. Note that the proxy dataset consists of two parts, i.e., a training part and a test part. Here, the cloud server only uses the training part to pre-train the global model.
    \item \textbf{Step 2: APoZ Score Calculation.} The cloud server inputs the test part of the proxy dataset into the pre-trained global model and records the activation of each neuron. Then the could server calculates the APoZ scores of each layer based on the activation of its neurons.
    \item \textbf{Step 3: Heterogeneous Model Generation.} The cloud server first specifies multiple levels of heterogeneous models with varying parameter sizes.
    For example, the cloud server can specify to generate three levels of heterogeneous models with 25\% (small), 50\% (medium), and 100\% (large) parameters of the original global model, respectively.
    For each heterogeneous model, the cloud server calculates the pruning ratio for each layer according to its APoZ score,  adjustment weight, and target model pruning ratio.
    The cloud server then generates multiple heterogeneous models by pruning the global model according to its calculated pruning ratio.
    Note that our pruning strategy ensures that a small model is a submodel of any model larger than it.
    The generated heterogeneous models are stored in the model pool.
    
\end{itemize}

The FL training stage consists of multiple FL training rounds.
As shown in Figure~\ref{fig: framework}, the cloud server maintains a table to record the device resource configuration, which can be requested directly from each device.
In each FL training round, the cloud server selects multiple devices for local training according to their resources.
Then, the cloud server dispatches the heterogeneous models in the model pool to the selected AIoT devices.
Note that each device is dispatched with a model coupled with 
its APoZ scores to guide local pruning.
Here, the APoZ score is an array with the length of the number of global model layers, whose communication overhead is negligible.
The device adaptively prunes the model according to its currently available resources before conducting local training.
To improve the performance of the large model, the devices perform a self-KD-based training strategy, which uses the output of the small models to guide the training of the large model.
Note that since small models are pruned from the large model, devices can directly obtain small models from their dispatched model.
After local training, devices upload their trained model to the cloud server.
The cloud server aggregates all the local models to update the global model and uses the new global model to update the models in the model pool.
Specifically, as shown in Figure \ref{fig: framework}, the workflow of each FL training round consists of seven steps as follows:
\begin{itemize}
    \item \textbf{Step 1: Model \& Device Selection.} The cloud server selects devices to participate in local training and selects a model from the model pool for each device according to their resources. 
    \item \textbf{Step 2: Model Dispatching.} The cloud server dispatches models from the model pool to its corresponding selected devices for local training. 
    Note that the cloud server also sends the APoZ scores to devices for local pruning.
    \item \textbf{Step 3: Adaptive Local Pruning.} Due to various uncertain factors, the available resources of a device may not be sufficient to enable training of the dispatched model.
    To facilitate local training, a device prunes its received models
    when its available resources are insufficient. 
    FlexFL enables a device to prune partial parameters based on its received model.
    If the device cannot train the model, it will be pruned directly to a smaller model.
    \item \textbf{Step 4: Self-KD-based Local Training.} Each device uses its local data to train the pruned model.
    Since a small model is the subset of any larger model, the pruned model includes all the parameters of small models.
    Small models can be trained on more devices, which means that small models are more adequately trained.
    To improve the performance of large models, devices calculate the loss for large model training using the soft label of small models together with the true label of data samples.
    
    \item \textbf{Step 5: Model Uploading.} Each device uploads its trained model to the cloud server for aggregation.
    \item \textbf{Step 6: Model Aggregation.} The cloud server aggregates the corresponding parameters of local models to generate a new global model.
    \item \textbf{Step 7: Model Pool Updating.} The cloud server uses the new global model to update the parameters of each model in the model pool.
\end{itemize}

\subsection{APoZ-Guided Model Generation}
\label{sec: APoZ-Guided Model Generation}
FlexFL generates multiple heterogeneous models by pruning the global model.
To generate high-performance heterogeneous models for local training, FlexFL aims to assign a higher pruning ratio to layers with more redundant parameters.
Based on this motivation, FlexFL adopts the APoZ~\cite{hu2016network} score as a metric to guide the model generation.

\subsubsection{APoZ Score Calculation (APoZCal($\cdot$))}
APoZ is a metric that measures the number of neuron activations after the ReLU layer. For each ReLU layer $i$, APoZ is defined as:
\begin{equation}
\label{eq: APoZ}
    A_i= \frac{\sum_{j=1}^{|\mathcal{D}^{test}_{p}|}{\sum_{k=1}^{N}{f(h^i_k(\mathcal{S}_j)}= 0) }}{|\mathcal{D}^{test}_{p}| \times N},
\end{equation}
where $f(\delta)$ is a boolean function, which returns 1 when the bool statement $\delta\models \top$, $N$ denotes the dimension of output feature map after ReLU layer, $|\mathcal{D}^{test}_{p}|$ denotes the total size of the test part of the proxy dataset and $h^i_k(\mathcal{S}_j)$ denotes the $k^{th}$ output feature map of the $j^{th}$ sample $\mathcal{S}_j$ after $i^{th}$ ReLU layer.

For models consisting of multiple residual blocks, e.g., ResNet~\cite{ResNet} and MobileNet~\cite{mobilenetv2}, we adjust the number of channels between blocks.
When a block contains multiple ReLU layers, we average its APoZs as the APoZ score of this block.
Specifically, if $i^{th}$ block contains $K_i$ ReLU layers, the block APoZ is calculated as:

\begin{equation}
\label{eq: blockAPOZ}
A_i= \frac{1}{K_i}{\sum_{t=1}^{K_i}\dfrac{\sum_{j=1}^{|\mathcal{D}^{test}_{p}|}{\sum_{k=1}^{N}{f(h^t_k(\mathcal{S}_j)}= 0)}}{|\mathcal{D}^{test}_{p}| \times N}}.
\end{equation}



\subsubsection{Adjustment Weight Calculation (AdjWCal($\cdot$))}
Typically, model 
layers with more parameters often contain more redundant neurons, 
diminishing the significance of individual neurons within layers. 
When pruning a specific number of neurons, prioritizing layers with more parameters is likely to have less impact on overall model performance compared to layers with fewer parameters.
Therefore, we calculate the adjustment weight as follows to adjust APoZ scores:
%
\begin{equation}
\label{eq: AdjW}
\small
AdjWCal(l_i,M) = \frac{\log{\operatorname{size}(l_i)}}{\log{\max(\operatorname{size}(l_1),...,\operatorname{size}(l_{len(M)}))}},
\end{equation}
where $l_i$ is the $i^{th}$ layer of $M$, the function $\operatorname{size}(l_i)$ calculates the number of parameters of the $i^{th}$ layer, and $len(M)$ denotes the number of layers of the model $M$.


\subsubsection{Heterogeneous Model Generation (ModelGen($\cdot$))} 
\label{sec: Model Gen}

Based on the calculated APoZ scores and adjustment weights, FlexFL can prune the global model to generate heterogeneous models.
Note that the heterogeneous model generation is performed on the server side and only once upon initialization.
Algorithm \ref{algorithm: Generate} presents the process of heterogeneous model generation.
Lines~\ref{Al2:1}-~\ref{Al2:2} initialize model pool $P$ and pruning ratios $s$ for each target model.
Lines~\ref{Al2:gen_start}-~\ref{Al2:gen_end} generates the multiple models according to the target model pruning ratios in $L_p$.
Lines~\ref{Al2:init1}-~\ref{Al2:init3} initialize the pruning control variable $\gamma$, the target model size $p_i$, and the target model $M^\prime$, respectively.
Line~\ref{Al2:while_start} evaluates the gap between the size of $M^\prime$ and 
a target model size $p_i$. 
Line ~\ref{Al2:AdjW_j} uses Equation \ref{eq: AdjW} to calculate the adjustment weight $AdjW_j$ for layer $l_j$.
Line~\ref{Al2:8} generates pruning ratio $s[i][j]$ based on APoZ score $A_j$ and adjustment weight $AdjW_j$.
In Line~\ref{Al2:9}, we set the minimum pruning ratios of each layer to 0.01 to avoid the parameters of a layer being completely pruned.
In Line~\ref{Al2:11}, according to the pruning ratios $s[i][]$ generated, we prune the global model $M$ to $M^\prime$.
Specifically, for layer $l_j$, $y_j$ and $x_j$ represent the numbers of output and input channels, according to the pruning ratios \( s[i][] \), we prune it to a model \( M' \) with  \( x_j \times s[i][j-1] \) input channels and \( y_i \times s[i][j] \)  output channels.
Note that if $W_i \in \mathbb{R}^{y_j \times x_j}$ is the hidden weight matrix of the global model $M$ in layer $l_j$, after pruning, $W'_i \in \mathbb{R}^{(y_j \times s[i][j]) \times (x_j \times s[i][j-1])}$ is the new hidden weight matrix of the pruned model $M^\prime$ in layer $l_j$.
In Line~\ref{Al2:14}, for $M^\prime$ with an error less than or equal to $\epsilon$, we add $M^\prime$ to the model pool $P$ as the pruned model corresponding to the target model pruning ratio $L_p[i]$.

\begin{algorithm}[h]
        \caption{Heterogeneous Model Generation}
        \label{algorithm: Generate}
        \KwIn{i) $S_{APoZ}$, the set of APoZ scores for each layer; ii) $M$, global model, iii) $L_{p}$, list of target model pruning ratios.}
        \KwOut{$P$, the model pool.}
        $P \leftarrow \{\}$\label{Al2:1}\\ 
        $s[i][j] \leftarrow 1 $ for $i \in [1,len(L_p)]$, $j \in [1,len(S_{APoZ})]$\label{Al2:2}\\
        \For{$i=1,\ldots,len(L_p)$}
        {\label{Al2:gen_start}
            $\gamma \leftarrow 0$ \label{Al2:init1}\\
            $p_i \leftarrow L_p[i] \times \operatorname{size}(M)$\label{Al2:init2}\\
            $M^\prime\leftarrow M$\label{Al2:init3}\\
            \While{$|p_i-\operatorname{size}(M^{\prime})| > \epsilon$}
            {\label{Al2:while_start}
                \For{$j=1,\ldots,len(S_{APoZ})$} 
                    {\label{Al2:prune_start}
                        $\langle l_j, A_j\rangle \leftarrow S_{APoZ}[j]$\\
                        $ AdjW_j \leftarrow \operatorname{AdjWCal}(l_j,M)$\label{Al2:AdjW_j}\\
                        $s[i][j] \leftarrow (1 - A_j \times AdjW_j )\times \gamma$\label{Al2:8}\\
                        $s[i][j] \leftarrow \operatorname{max}(\operatorname{min}(s[i][j],1),0.01)$\label{Al2:9}\\
                    } \label{Al2:prune_end}
                $M^{\prime} \leftarrow \operatorname{prune}(M,s[i])$\label{Al2:11} \\
                $\gamma \leftarrow \gamma + \xi \qquad$//$\xi=0.01$ is the iteration step.\label{Al2:17}\\
            }
            $P \leftarrow P \cup \{M^{\prime}\}$\label{Al2:14}\\
        }\label{Al2:gen_end}
        \textbf{return} $P$
\end{algorithm}


\subsection{Adaptive Local Model Pruning (AdaPrune($\cdot$))}
\label{sec: AdaPrune}

To address the problem of insufficient
available resources within  uncertain scenarios, 
FlexFL enables devices to adaptively prune their received models to adaptive models for local training, focusing on memory resource constraints.
Specifically, when a device does not have sufficient memory resources for training its received model, it first prunes $\Gamma\times size(M)$ parameters to generate an adaptive model for local training, where $\Gamma$ is the adaptive pruning size and $size(M)$ is the number of parameters of the global model $M$.
Note that our approach requires
that the size of the pruned parameters (i.e., $\Gamma\times size(M)$) 
should be smaller than the smallest size of the parameter differences between any two models. 
When its available resources are still insufficient to train the pruned model, the device directly prunes it to a smaller model that best fits the device. 
For example, assume that $M_1$, $M_2$, and $M_3$ denote the small, medium, and large models, respectively. Let   $M_2^{\prime}$ and $M_3^\prime$ be the adaptive models of $M_2$ and $M_3$, respectively. 
If a model of type $M_3$ is dispatched to some device with insufficient resources, the client will adaptively prune the model in the order of  $M^\prime_3$, $M_2$,  $M^\prime_2$, and $M_1$ until the pruned model can be accommodated by the device.

Similar to Algorithm \ref{algorithm: Generate},  the adaptive pruning is still based on our calculated APoZ scores.
Since our APoZ scores are calculated before FL training and are not updated within the training process, the cloud server can directly send the APoZ scores to all devices.
In addition, since all heterogeneous models in FlexFL are pruned from the same global model without any extra exit layers, and a smaller model is a submodel of a larger model, devices can prune the received model according to the corresponding pruning scheme.

\subsection{Self-Knowledge Distillation-based Local Training}
\label{sec: Self-Knowledge Distillation-based Local Training}
In resource-constrained scenarios, the majority of devices cannot train large models, which results in inadequate training for large models.
To improve the performance of large models, FlexFL adopts a Knowledge Distillation~\cite{hinton2015distilling,phuong2019towards} strategy.
Since small models are the submodel of large models, devices can utilize adequately trained small models to enhance the training of large training.
Specifically, during model training, the device can obtain the outputs (i.e., soft labels) of the large model and small models. Then the device calculates the Cross-Entropy (CE) loss using the output of the large model and true labels and calculates the Kullback-Leibler (KL) loss based on the outputs of the large model and small models.
Finally, the device uses both CE and KL losses to update the model.
Assume that $M_i$ is a large model, $\hat{y}_i$ is the soft labels of model $M_i$ and ${y}_i$ is ground truth, the CE loss of $M_i$ is defined as:
$$\mathcal{L}_{CE} = - \log {h(\hat{y}_i)}[y_i],$$
where $h(\cdot)$ is the softmax function.
Assume that $M_1$, $M_2$, ..., $M_{i-1}$ are the smaller models for $M_i$ and $\hat{y}_{1}$, $\hat{y}_{2}$, ..., $\hat{y}_{i-1}$ are the soft labels of these models.
The KL loss can be calculated as follows:
$$\mathcal{L}_{KL} = \frac{1}{i-1}{\sum_{j=1}^{i-1}\operatorname{sum}{(h(\hat{y}_{j})/\tau)}\cdot \tau^2 \log \frac{h(\hat{y}_{j})}{h(\hat{y}_{i})}},$$
where $\tau$ is the temperature to control the distillation process.
Note that a higher value of $\tau$ leads to smoother probability distributions, making the model focus more on relatively difficult samples, and a lower value of $\tau$ makes the probability distribution sharper, making the model more confident and prone to overfitting.




According to the CE loss and the KL loss, we can obtain the final loss $\mathcal{L}$ of the model $M_i$ as follows:
\begin{equation}
\label{equ: L}
    \mathcal{L} = \mathcal{L}_{CE} + \lambda \mathcal{L}_{KL}
\end{equation}
where $\lambda$ is a hyperparameter that controls the training preference for two types of losses. 
Note that a large value of $\lambda$ makes the model training more influenced by KL divergence loss.
On the contrary, a small value of $\lambda$ guides the training to focus more on cross-entropy loss.

\subsection{Heterogeneous Model Aggregation (Aggr($\cdot$))}
When all the local models are received and saved in $S_{upload}$, 
 the cloud server can perform the aggregation. 
Since all the heterogeneous models are pruned from the global model, the cloud server generates a new global model by
aggregating the corresponding parameters of the models in $S_{upload}$ with the weights determined by the number of its trained data.

Assume that $p$ is a parameter in the global model $M$, 
and $\sigma(p, S_{upload})$ extracts the model in the set that contains the corresponding parameter of $p$ and the numbers of their training data.
Let $\theta$ be the parameters of the aggregated global model $M$, which can be calculated as follows:
$$\forall{p\in \theta},\; p = \frac{\sum_{m\in \sigma(p, S_{upload})} p^m\times d^m}{\sum_{m\in \sigma(p, S_{upload})} d^m},$$
where $p^m$ denotes the corresponding parameter of $p$ in $m$ and $d^m$ is the number of training data of $m$.


By applying the above equation to aggregate all the parameters of models in $S_{upload}$, the cloud server can generate an updated global model.
Subsequently, the cloud server updates all the heterogeneous models in the model pool $P$ by assigning the parameter values of the aggregated global model to their corresponding parameters.

\subsection{Implementation of FlexFL}

{
\color{red}
\begin{algorithm}[h]
        \caption{Implementation of FlexFL}
        \label{algorithm: Main}
        \KwIn{i) $T$, training rounds;
        ii) $D$, the set of devices; 
        iii)$f$, fraction of selected devices;
        iv) $M$, global model;
        v)  $\mathcal{D}_p$, proxy dataset, vi) $L_{p}$, list of pruning ratios of heterogeneous models.}
        $T_r \leftarrow \operatorname{ResConfRequest}(D)$ \label{line:pre_start} \\
        $S_{APoZ} \leftarrow \operatorname{APoZCal} (M, \mathcal{D}_{p})$ \label{line:pretrain} \\
        $ \operatorname{Reset}(M)$ \label{line:reset} \\
        $\{M_1,M_2,...,M_{p}\}\leftarrow \operatorname{ModelGen}(S_{APoZ},M, L_p)$ \label{line:model_gen} \\
        $P\leftarrow \{M_1,M_2,...,M_{p}\}$\label{line:pre_end} \\
        \For{epoch $ e = 1,\ldots,T$}{\label{line:fl_train_start}
            $K \leftarrow \operatorname{max}(1, f|D|) $ \label{line:K} \\ 
            $S_d \leftarrow \operatorname{DevSel}(D, P, K)$  \label{line:S_d}\\
            $S_{upload}\leftarrow \{\}$ \label{line:local}\\
            /*parallel for*/\\
            \For{$\langle d_k, m_k \rangle$ in $S_d$ }{\label{line:local_train_start}
                $r_{d_k} \leftarrow \operatorname{ResRequest}(d_k)$ \label{line:r_ck}\\
                $m'_k \leftarrow \operatorname{AdaPrune} \left(r_{d_k}, m_k\right)$ \label{line:m'k}\\
                $L \leftarrow \mathcal{L}(m^\prime_k,\mathcal{D}_k)$ \label{line:L}\\
                $\theta^\prime_{m_k} \leftarrow \theta^\prime_{m_k} - \frac{\partial{L}}{\partial{\theta^\prime_{m_k}}}$ \label{line:theta}\\
                $S_{upload}\leftarrow S_{upload}\cup \{\langle m_k^{\prime}, \left|\mathcal{D}_k\right| \rangle\}$\label{line:b_local}\\
                }\label{line:local_train_end}
            $M \leftarrow \operatorname{Aggr}(S_{upload})$\label{line:M}\\
            $P \leftarrow \operatorname{update}(P,M)$\label{line:model_update}\label{line:P}
            }\label{line:fl_train_end}
\end{algorithm}}

Algorithm \ref{algorithm: Main} presents the implementation of our FlexFL approach. 
Lines~\ref{line:pre_start}-\ref{line:pre_end} denote the operations of pre-processing.
Line~\ref{line:pre_start} initializes the resource configuration table, where function \textit{ResConfRequest($\cdot$)} requests all the devices to upload their available resource information.
In Line~\ref{line:pretrain}, the function \textit{APoZCalculate($\cdot$)} pretrains the global model $M$ using the training part of the proxy dataset $\mathcal{D}^{train}_{p}$ and calculates the APoZ scores for each layer of $M$ using the test part of the proxy dataset $\mathcal{D}^{test}_{p}$, where $S_{APoZ}$ is a set of two-tuples $\langle l_i, A_i \rangle$, $l_i$ denotes the $i^{th}$ layer of $M$, $A_i$ denotes the APoZ score of the $i^{th}$ layer.
Line~\ref{line:reset} resets the global model.
Line~\ref{line:model_gen} generates $len(L_p)$ heterogeneous models according to calculated APoZ scores.
Line~\ref{line:pre_end} stores the generated models in the model pool $P$.
Lines~\ref{line:fl_train_start}-\ref{line:fl_train_end} present the process of FL training stage.
Line~\ref{line:K} calculates the number of devices needed to participate in local training.
In Line~\ref{line:S_d}, the function \textit{DevSel($\cdot$)} selects $K$ devices and their respective trained models, where $S_d$ is a set of two tuples $\langle d, m\rangle$, $d\in D$ is a selected device, and $m\in P$ is a model that will be dispatched to $d$.
Line~\ref{line:local} initializes the model set $S_{upload}$, a set of two tuples $\langle m, num \rangle$, where $m$ is a local model and $num$ is the number of data samples used to train $m$.
Lines~\ref{line:local_train_start}-\ref{line:local_train_end} present the local training process.
Line~\ref{line:r_ck} requests the current resources of $d_k$ and Line~\ref{line:m'k} uses our adaptive local pruning strategy to prune the received model $m_k$ according to $r_{d_k}$.
Line~\ref{line:L} employs Equation \ref{equ: L} to calculate the loss $L$ and Line~\ref{line:theta} updates the parameters of $m^\prime_k$ according to $L$, where $\theta^\prime_{m_k}$ denotes the parameters of $m^\prime_k$.
In Line~\ref{line:b_local}, the device uploads its trained model $m^\prime_k$ together with the number of data samples $\left|\mathcal{D}_k\right|$ to the model set $S_{upload}$.
Line~\ref{line:M} aggregates all the models in $S_{upload}$ to update the global model $M$.
Line~\ref{line:model_update} updates the models in $P$ using the global model $M$.

%% file: experiment.tex
\section{Performance Evaluation}\label{section: experi}
To evaluate FlexFL performance, we implemented FlexFL using  PyTorch.
For all investigated FL methods, we adopted the same SGD optimizer with a learning rate of 0.01 and a momentum of 0.5. For local training, we set the batch size to 50 and the local epoch to 5.
We assumed that $|D|=100$ AIoT devices were involved in total, and $ f=10\%$ of them were selected in each FL training round. All the experiments were conducted on an Ubuntu workstation with one Intel i9 13900k CPU, 64GB memory, and one NVIDIA RTX 4090 GPU.

\subsection{Experimental Settings}
\subsubsection{Device Heterogeneity Settings}\label{subsubsec: Device Heterogeneity Settings}
\label{sec: device settings}
To evaluate the performance of FlexFL in uncertain and resource-constrained scenarios, we simulated various devices with different dynamic resources (available memory size).
Specifically, we employed Gaussian distribution to define dynamic device resources as follows: $r = r_M - |u|$, where $r_M$ is the maximum memory capacity of the device and $u \sim \mathcal{N}(0,\sigma^2)$.
\begin{table}[h]
\vspace{-0.15in}
\centering
\caption{Device uncertainty settings.}
\vspace{-0.1in}
\scriptsize
\label{Table: uncertain devices}
\begin{tabular}{cccc}
\hline
Level   & \# Device & Maximum Capacity $r_M$ & Variance $u$            \\ \hline
Weak  & 40\%      & $r_M = 35$             & $\sigma^2 \in [5,8,10]$ \\
Medium & 30\%      & $r_M = 60$             & $\sigma^2 \in [5,8,10]$ \\
Strong  & 30\%      & $r_M = 110$            & $\sigma^2 \in [5,8,10]$ \\ \hline
\end{tabular}
\vspace{-0.05in}
\end{table}

In our experiment, we adopted three levels of devices, i.e., weak, medium, and strong. We set the ratio of the number of devices at these three levels to 40\%, 30\%, and 30\%, respectively.
The distribution of their available memory size across these devices is uncertain, as shown in Table \ref{Table: uncertain devices}.
If the device memory is smaller than its received model $m$, i.e., $r \leq \frac{\text{size}(m)}{\text{size}(M)}\times 100$, our approach will not train $m$ due to insufficient memory resources. 
In this case,  $m$ will be pruned to be an adaptive model to fit the device. For example, if $r$ is 30, the number of parameters of a pruned model cannot exceed   30\%  of its original counterpart.

\begin{table*}[!h]
\centering
\caption{Test accuracy (\%) of average and large models. The best results are shown in \textbf{bold}.}
\vspace{-0.1in}
\label{Table: Experiment}
\scriptsize
\begin{tabular}{cc|ccc|ccc|ccc}
\hline
\multirow{2}{*}{Model}       & \multirow{2}{*}{Algorithm} & \multicolumn{3}{c|}{CIFAR10}                                       & \multicolumn{3}{c|}{CIFAR100}                                      & \multicolumn{3}{c}{TinyImagenet}                                   \\ \cline{3-11} 
                             &                            & IID                  & $\alpha = 0.6$       & $\alpha = 0.3$       & IID                  & $\alpha = 0.6$       & $\alpha = 0.3$       & IID                  & $\alpha = 0.6$       & $\alpha = 0.3$       \\ \hline
\multirow{4}{*}{VGG16}       & Decoupled                  & 75.81/72.76          & 73.57/69.91          & 69.95/66.01          & 34.53/29.44          & 33.68/29.12          & 33.33/28.77          & 25.62/21.14          & 26.56/22.53          & 26.70/23.51          \\
                             & HeteroFL                   & 79.75/76.92          & 77.53/75.45          & 73.89/71.50          & 34.37/29.92          & 35.02/31.31          & 35.43/31.56          & 25.51/23.19          & 26.63/25.01          & 27.90/26.78          \\
                             & ScaleFL                    & 80.36/77.72          & 75.99/74.00          & 72.95/69.87          & 34.10/29.00          & 34.26/29.38          & 33.60/29.18          & 22.37/19.38          & 23.15/20.96          & 24.67/21.62          \\
                             & FlexFL                     & \textbf{84.75/85.16} & \textbf{83.11/83.45} & \textbf{80.60/81.06} & \textbf{40.27/40.35} & \textbf{41.29/41.30} & \textbf{39.55/39.99} & \textbf{26.61/26.97} & \textbf{29.19/29.56} & \textbf{31.07/31.42} \\ \hline
\multirow{4}{*}{Resnet34}    & Decoupled                  & 62.92/56.97          & 59.01/54.11          & 55.35/51.32          & 26.88/22.28          & 26.68/22.01          & 25.24/19.81          & 33.13/29.07          & 32.95/25.03          & 31.33 24.80          \\
                             & HeteroFL                   & 69.56/63.17          & 65.14/61.25          & 61.29/57.02          & 31.79/24.52          & 31.01/25.37          & 30.92/24.68          & 38.35/31.90          & 36.52/33.19          & 35.02/32.67          \\
                             & ScaleFL                    & 76.65/74.23          & 71.57/68.50          & 66.56/61.67          & 36.84/31.87          & 34.95/29.51          & 32.81/27.50          & 38.31/32.84          & 36.68/31.32          & 36.02/31.77          \\
                             & FlexFL                     & \textbf{77.48/78.06} & \textbf{72.89/73.71} & \textbf{69.08/69.60} & \textbf{37.76/37.38} & \textbf{37.63/37.31} & \textbf{37.34/37.38} & \textbf{40.53/40.85} & \textbf{38.64/38.32} & \textbf{37.89/37.88} \\ \hline
\multirow{4}{*}{MobileNetV2} & Decoupled                  & 53.01/52.03          & 48.34/47.71          & 42.42/40.29          & 20.20/17.74          & 20.74/18.08          & 20.22/17.58          & 21.57/16.11          & 20.34/17.60          & 20.11 16.35          \\
                             & HeteroFL                   & 57.60/52.69          & 51.31/48.33          & 44.00/40.16          & 23.49/19.31          & 22.60/19.20          & 21.60/17.80          & 24.96/20.05          & 24.96/22.23          & 22.52/20.82          \\
                             & ScaleFL                    & 63.42/59.83          & 54.57/48.77          & 49.90/45.10          & 26.53/21.72          & 25.09/19.90          & 24.30/18.24          & 26.64/23.63          & 26.31/23.43          & 24.69/22.25          \\
                             & FlexFL                     & \textbf{68.47/69.18} & \textbf{60.00/61.32} & \textbf{56.87/58.23} & \textbf{29.11/28.30} & \textbf{27.86/27.14} & \textbf{26.78/26.41} & \textbf{28.26/27.44} & \textbf{27.46/25.55} & \textbf{25.38/24.00} \\ \hline
\end{tabular}
\vspace{-0.15in}
\end{table*}

\subsubsection{Data Settings}
In our experiments, we utilized three well-known datasets, i.e., CIFAR-10~\cite{CIFAR}, CIFAR-100~\cite{CIFAR}, and TinyImagenet~\cite{le2015tiny}. To investigate the performance on non-IID scenarios, we adopted the Dirichlet distribution $Dir(\alpha)$ to assign data to the devices involved. By controlling the hyperparameter $\alpha$ of the Dirichlet distribution, we managed the degree of IID bias in the data, where smaller values of $\alpha$ indicate higher data heterogeneity.

\subsubsection{Model Settings}
To validate the generality of our method, we conducted experiments under the models of different sizes and different architectures, i.e., VGG16~\cite{VGG}, ResNet34~\cite{ResNet}, and MobileNetV2~\cite{mobilenetv2}.

\begin{figure}[h]
\vspace{-0.15in}
\centering
    \subfloat[\scriptsize FlexFL]{\includegraphics[width=0.16\textwidth]{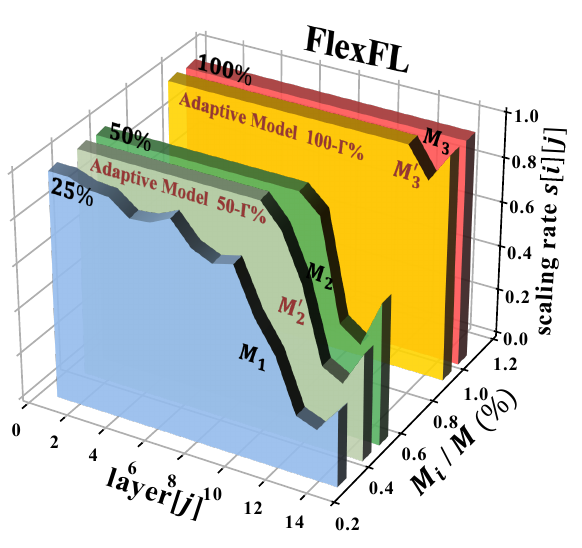}%
    \label{fig:flexfl}}
\hspace{-0.05in}
    \subfloat[\scriptsize ScaleFL]{\includegraphics[width=0.16\textwidth]{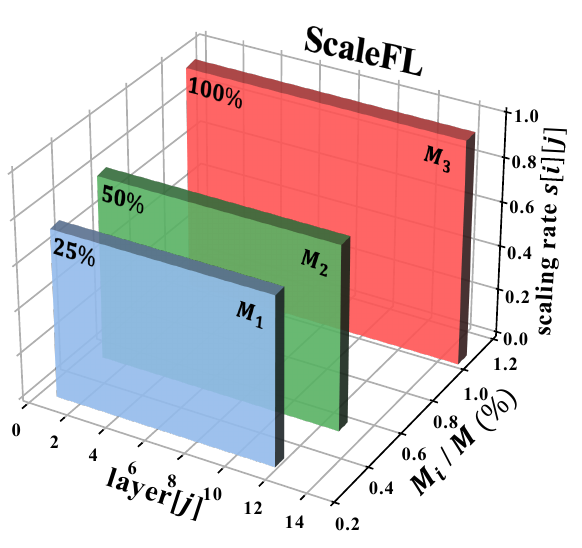}%
    \label{fig:scalefl}}
 \hspace{-0.05in}
    \subfloat[\scriptsize HeteroFL]{\includegraphics[width=0.16\textwidth]{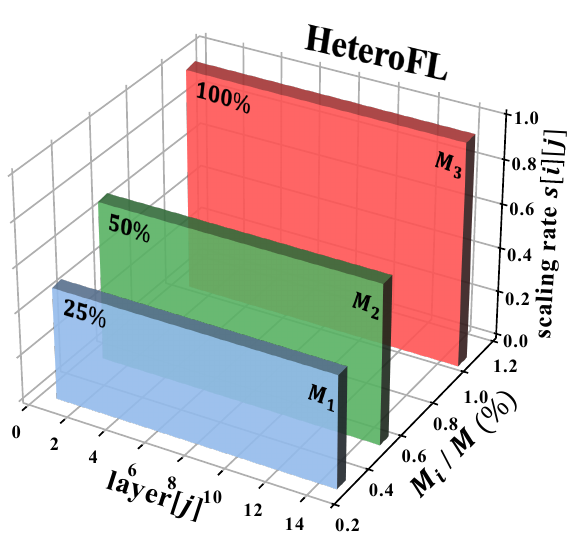}%
    \label{fig:heterofl}}
\caption{Comparison of submodels  (VGG16 on CIFAR10).}
\label{fig: Comparison of submodels in different heterogeneous FL}
\vspace{-0.25in}
\end{figure}

We adopted $p=3$ and uniformly set the list of target model pruning ratios $L_p$ to [25\%, 50\%, 100\%] for all methods, yielding a model pool $P =\{M_1, M_2, M_3\}$ with three models, where model $M_3$ is the largest model and also the global model.
Figure~\ref{fig: Comparison of submodels in different heterogeneous FL} presents an example to visualize the pruned models based on different FL methods. 
In the case of FlexFL, the adaptive models are $M^\prime_2$ and $M^\prime_3$, which are obtained by pruning $\Gamma \times size(M_3)$ parameters from $M_2$ and $M_3$, respectively, where $\Gamma = 10\%$. For self-KD hyperparameters, we set $\tau =3$ and $\lambda = 10$. 

\subsection{Performance Comparison}
\begin{figure}[h]
\vspace{-0.25in}
\centering
    \subfloat[\footnotesize CIFAR10, $\mathrm{IID}$]{\includegraphics[width=0.2\textwidth]{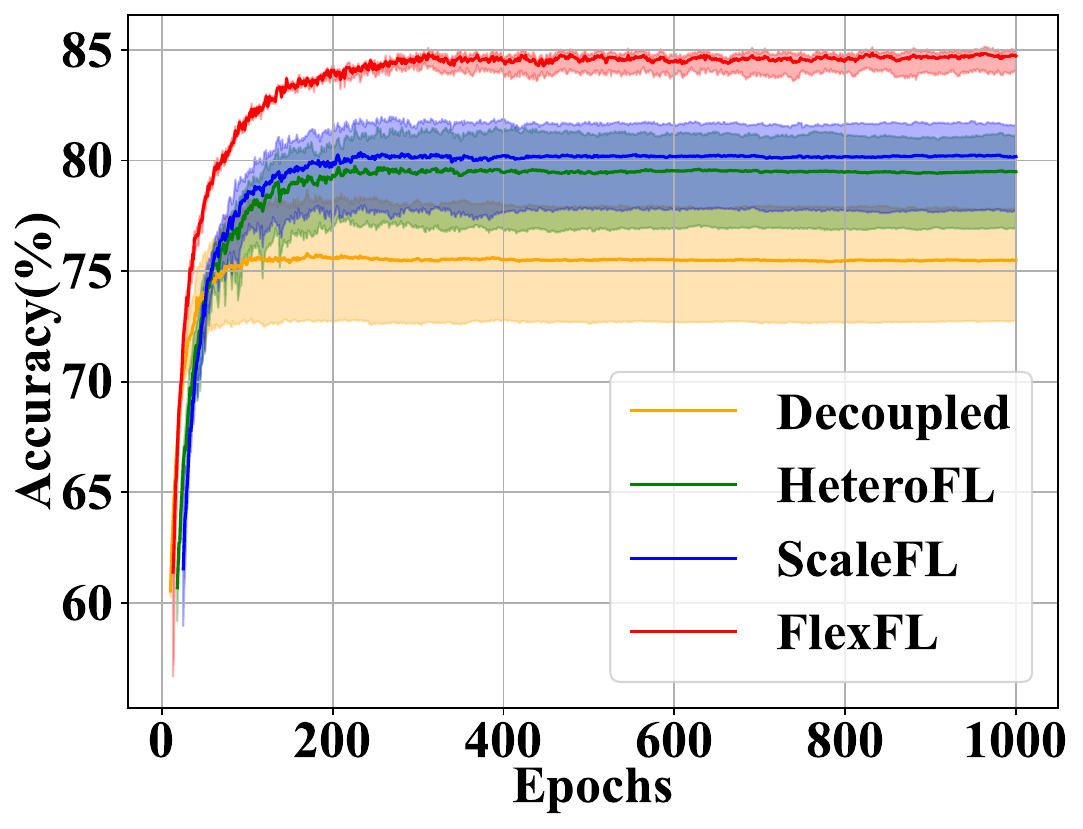}%
    \label{fig:vgg-cifar10-iid}}
    \hfil
    \subfloat[\footnotesize CIFAR10, $\alpha=0.3$]{\includegraphics[width=0.2\textwidth]{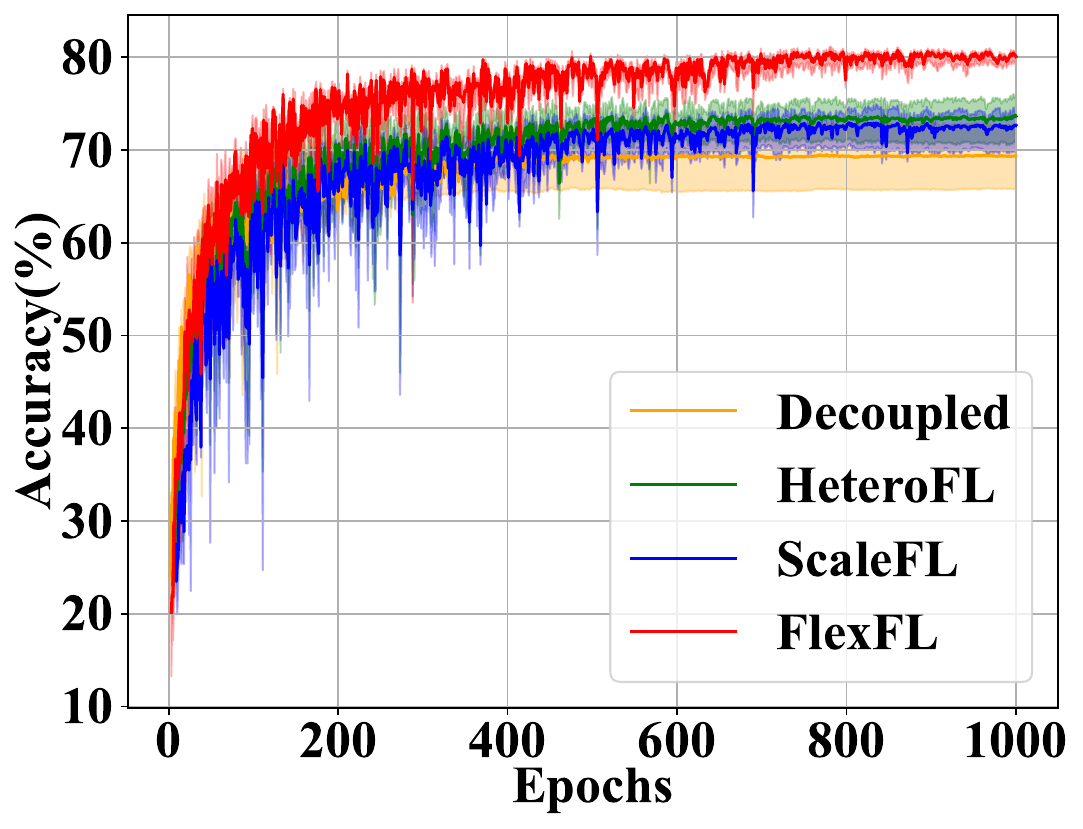}%
    \label{fig:vgg-cifar10-niid03}}
    \vspace{-0.1in}
    \subfloat[\footnotesize CIFAR100, $\mathrm{IID}$]{\includegraphics[width=0.2\textwidth]{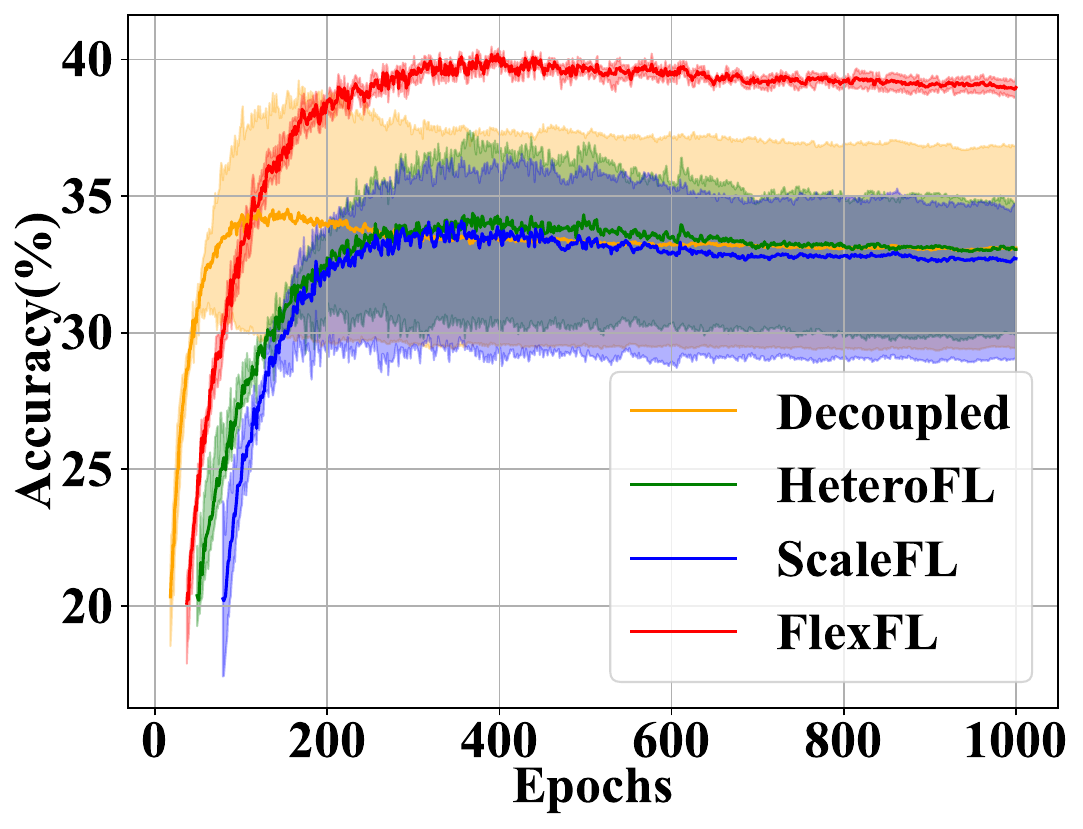}%
    \label{fig:vgg-cifar100-iid}}
    \hfil
    \subfloat[\footnotesize CIFAR100, $\alpha=0.3$]{\includegraphics[width=0.2\textwidth]{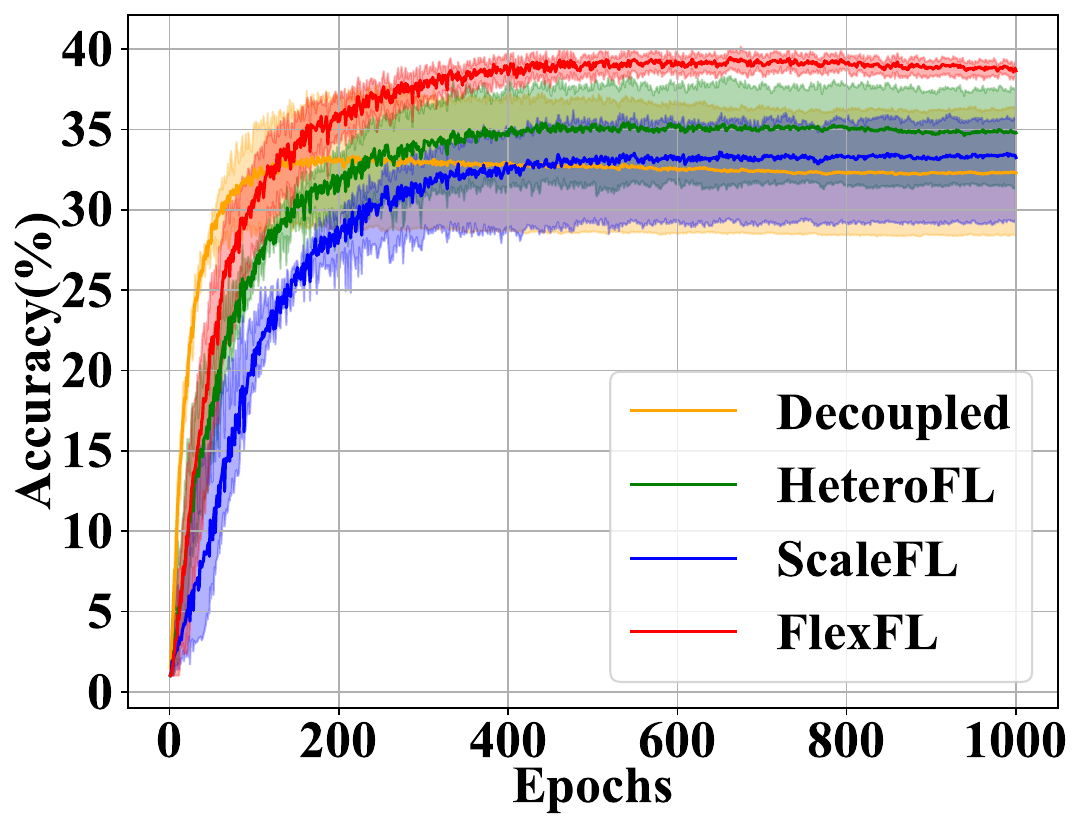}%
    \label{fig:vgg-cifar100-niid03}}
\caption{Learning curves of FlexFL and three baselines.}
\label{fig:comparison-for-different-methods}
\vspace{-0.05in}
\end{figure}

In our experiment, we compared three methods to our method: Decoupled\cite{FedAvg}, HeteroFL\cite{heterofl}, and ScaleFL\cite{ScaleFL}. 
Decoupled follows a strategy similar to FedAvg\cite{FedAvg}, where large, medium, and small models are trained on devices capable of hosting them without considering model aggregation.
HeteroFL generates corresponding models based on width-wise pruning.
ScaleFL, on the other hand, prunes models based on both width and depth proportions to create their corresponding models. 
Figure \ref{fig:comparison-for-different-methods} illustrates the comparison between the accuracy of our method and three other baselines.
The solid line in the middle represents the average accuracy of the 25\%, 50\%, and 100\% models, and the boundaries filled with corresponding colors represent the highest and lowest accuracies among all models.

\subsubsection{Large Model Performance Analysis}
In Table \ref{Table: Experiment}, we use the notation ``$x/y$''  to specify the test accuracy, where $x$ denotes the average accuracy of the 25\%, 50\%, and 100\% models and $y$ indicates the accuracy of the 100\% model. It is evident that FlexFL consistently achieves an accuracy improvement ranging from 1.75\% to 13.13\% in terms of large model accuracy, regardless of whether in the IID or non-IID scenarios. 
This indicates that our method performs better in obtaining high accuracy on larger models.
This discrepancy can be attributed to the greater flexibility in pruning offered by VGG16 and MobileNetV2. Specifically, VGG16 permits pruning of up to the first 15 layers, whereas MobileNetV2 allows for pruning of up to 9 blocks. In contrast, due to the inability to disrupt the internal structure of residual blocks in ResNet34, we can only prune 5 blocks, resulting in a relatively smaller improvement compared to the baselines.

\subsubsection{Average Model Performance Analysis}
Our experiment also evaluated the average model accuracy, as shown in Table \ref{Table: Experiment}.
Based on our observations of the datasets, our approach demonstrates an accuracy improvement ranging from 0.92\% to 7.65\% compared to ScaleFL.
We also observe that in most datasets, the accuracy of the largest model is higher than the average model.
Conversely, in the ScaleFL, HeteroFL, and Decoupled approaches, the accuracy of the largest model is lower than that of the average model. This indicates that in our approach, large models can effectively leverage their greater number of parameters, while in other methods, large models exhibit a paradoxical scenario where they possess more parameters but lower accuracy compared to smaller or medium-sized models.
This phenomenon arises from the fact that small models can be trained on all devices, whereas large models can only be trained on devices with ample resources. Consequently, small models encapsulate a broader spectrum of knowledge. In our approach, by distilling knowledge from large models to smaller ones, the larger models can enhance accuracy by assimilating knowledge from other models.

\subsection{Impacts of Different Configurations}
\subsubsection{Proxy Dataset Size}
\begin{figure}[h]
\vspace{-0.05in}
\centering
\includegraphics[width=0.28\textwidth]{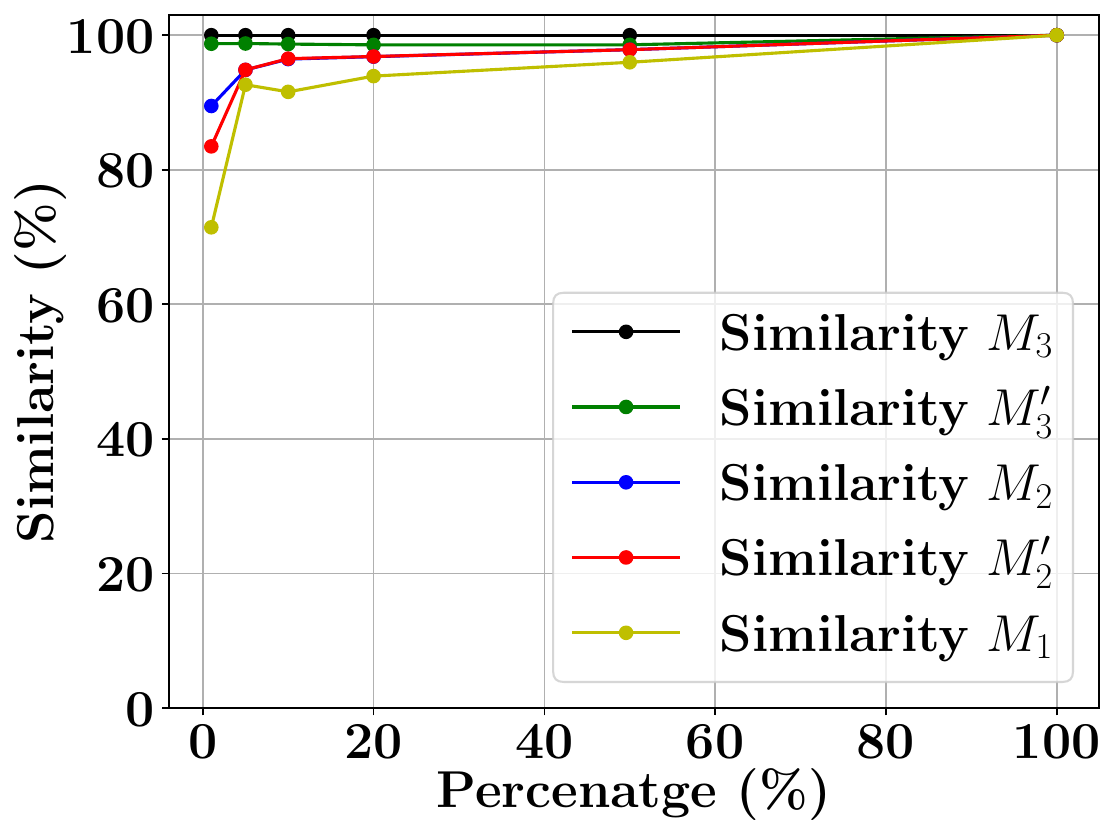}
\vspace{-0.05in}
\caption{Model pruning ratios similarity with different proxy dataset sizes.}
\vspace{-0.05in}
\label{fig: sim}
\end{figure}

To evaluate the impact of the pruning ratio $s$ on different proxy dataset sizes, we used training datasets of sizes 100\%, 50\%, 20\%, 10\%, 5\%, and 1\% as proxy datasets, where 80\% of the proxy dataset was used as the training set for pre-training. After training for 100 rounds, the remaining portion was used as the test set to calculate APoZ scores. Based on Algorithm \ref{algorithm: Generate}, we compared the similarity of pruning ratio $s_{p\%}[i]$ obtained from model $M_i$ in the model pool $P$, where $p\%$ represents the proxy dataset size.  The similarity is defined as:
$$sim_{(p\%,M_i)} = 1-\operatorname{avg}(\frac{|s_{p\%}[i]-s_{100\%}[i]|}{s_{100\%}[i]}).$$
Figure \ref{fig: sim} shows
the similarity for each level model $M_i$.
We can observe that even with only 1\% of the data, FlexFL achieves pruning ratios similar to those using the full dataset.

We conducted heterogeneous FL training using different model pools $P$ generated by different proxy dataset sizes.
As shown in Table~\ref{Table: similarity}, FlexFL achieves inference accuracy similar to that of the full dataset when using only 1\% data as a proxy dataset. 
Therefore, FlexFL can achieve good performance only by using a very small proxy dataset.

\begin{table}[h]
\vspace{-0.15in}
\centering
\caption{Test accuracy (\%) in different proxy dataset size.}
\vspace{-0.1in}
\scriptsize
\label{Table: similarity}
\begin{tabular}{ccc}
\hline
\multicolumn{1}{c}{Proxy Dataset Size} & Global Model Accuracy  & Avg Model Accuracy \\ \hline
1\%                                    & 84.90\%           & 85.19\%              \\
5\%                                    & 84.37\%           & 84.63\%              \\
10\%                                   & 84.11\%           & 84.31\%              \\
20\%                                   & 84.60\%           & 84.88\%              \\
50\%                                   & 84.02\%           & 84.37\%              \\
100\%                                  & 84.56\%           & 84.90\%              \\ \hline
\end{tabular}
\vspace{-0.15in}
\end{table}



\subsubsection{Numbers of Involved Devices} 
\label{sec: Participating}
\begin{figure}[h]
\vspace{-0.2in}
\centering
    \subfloat[\footnotesize$|D|=50$]{\includegraphics[width=0.2\textwidth]{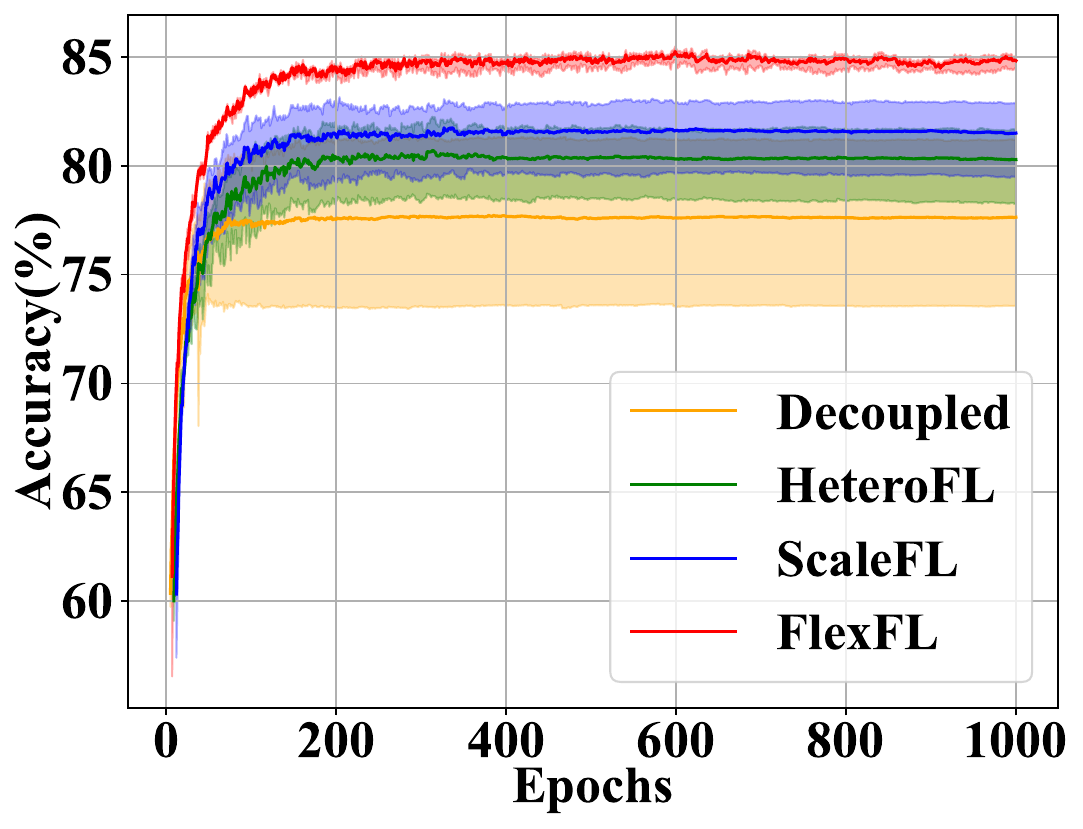}%
    }
    \hfil
    \subfloat[\footnotesize$|D|=100$]{\includegraphics[width=0.2\textwidth]{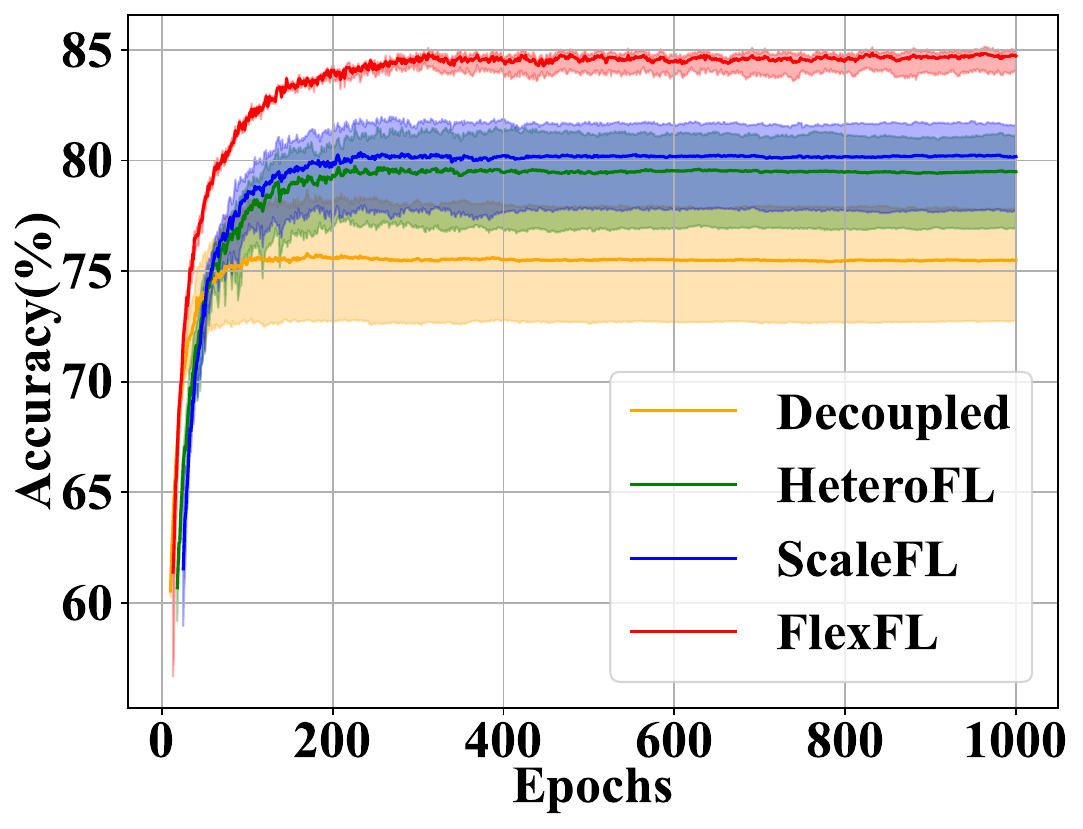}%
    }
    \vspace{-0.1in}
    \subfloat[\footnotesize$|D|=200$]{\includegraphics[width=0.2\textwidth]{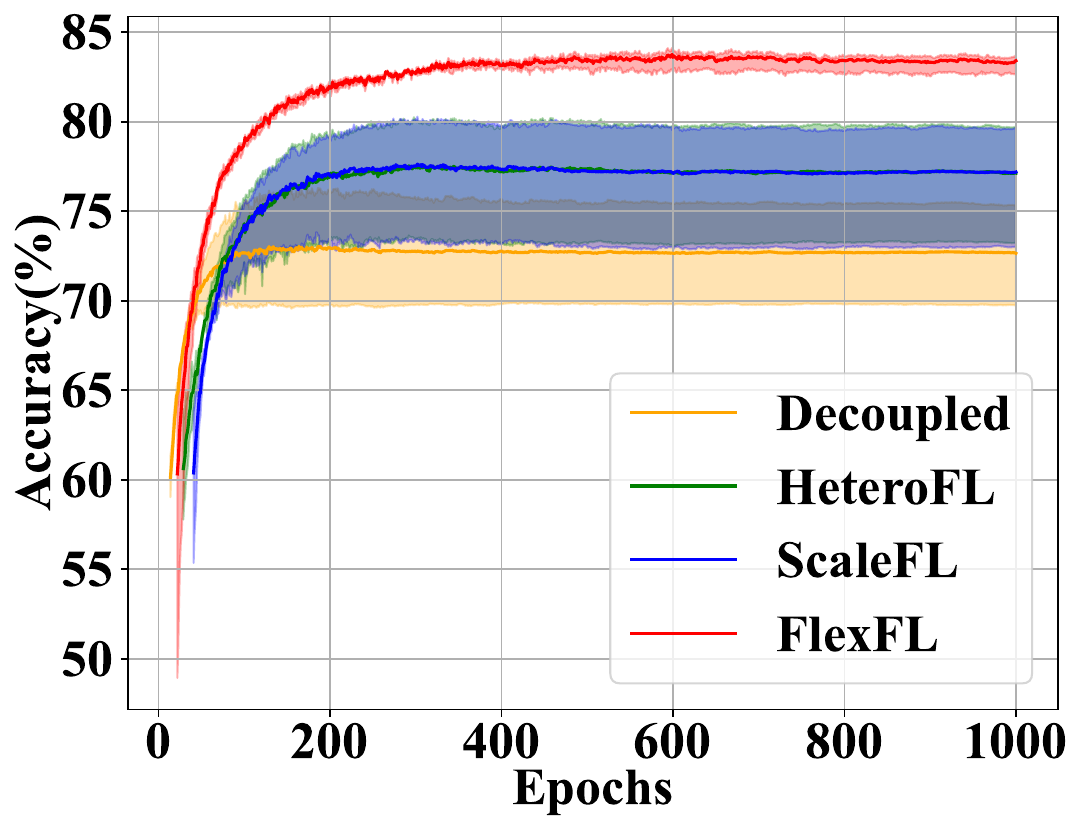}%
    }
    \hfil
    \subfloat[\footnotesize$|D|=500$]{\includegraphics[width=0.2\textwidth]{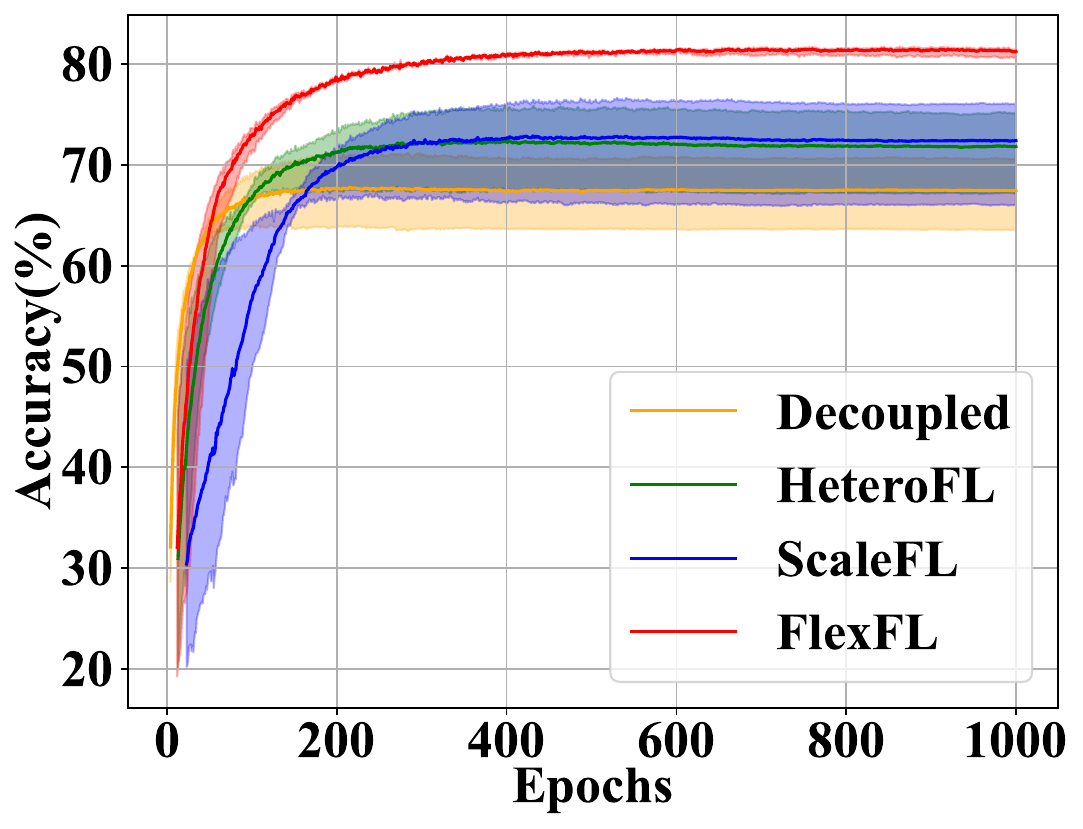}%
    }
 \caption{Learning curves for different numbers of involved devices.}
\label{fig:comparison-number of users}
\vspace{-0.15in}
\end{figure}

To investigate the scalability of our approach in various 
heterogeneous FL scenarios, 
we studied the impact of varying numbers of involved devices on inference accuracy. 
Specifically, we conducted experiments based on  CIFAR10 and VGG16 within  IID scenarios involving  $|D|=50$, $100$, $200$, and $500$ devices, respectively.
%
In each training round, 10\% of the devices were selected. 
Figure $\ref{fig:comparison-number of users}$ shows that our method consistently improves performance across different numbers of devices. Moreover, as $|D|$ increases, the accuracy of all methods decreases. 
The increase in the number of involved devices leads to less local data on each device, which reduces the generalization of the local model and further reduces the accuracy of the global model.
Specifically, our method experiences a 3.72\% decrease in average model accuracy when increasing from $|D|=50$ to $|D|=500$. In contrast, ScaleFL experiences an 8.86\% decrease in average model accuracy as the number of involved devices increases. 
This indicates that our method is more practical and adaptable to heterogeneous FL scenarios with large numbers of devices.

\subsubsection{Numbers of Selected Devices} 
\label{sec: Simultaneously}
\begin{figure}[h]
   \vspace{-0.2in}
\centering
    \subfloat[\footnotesize$f=5\%$]{\includegraphics[width=0.2\textwidth]{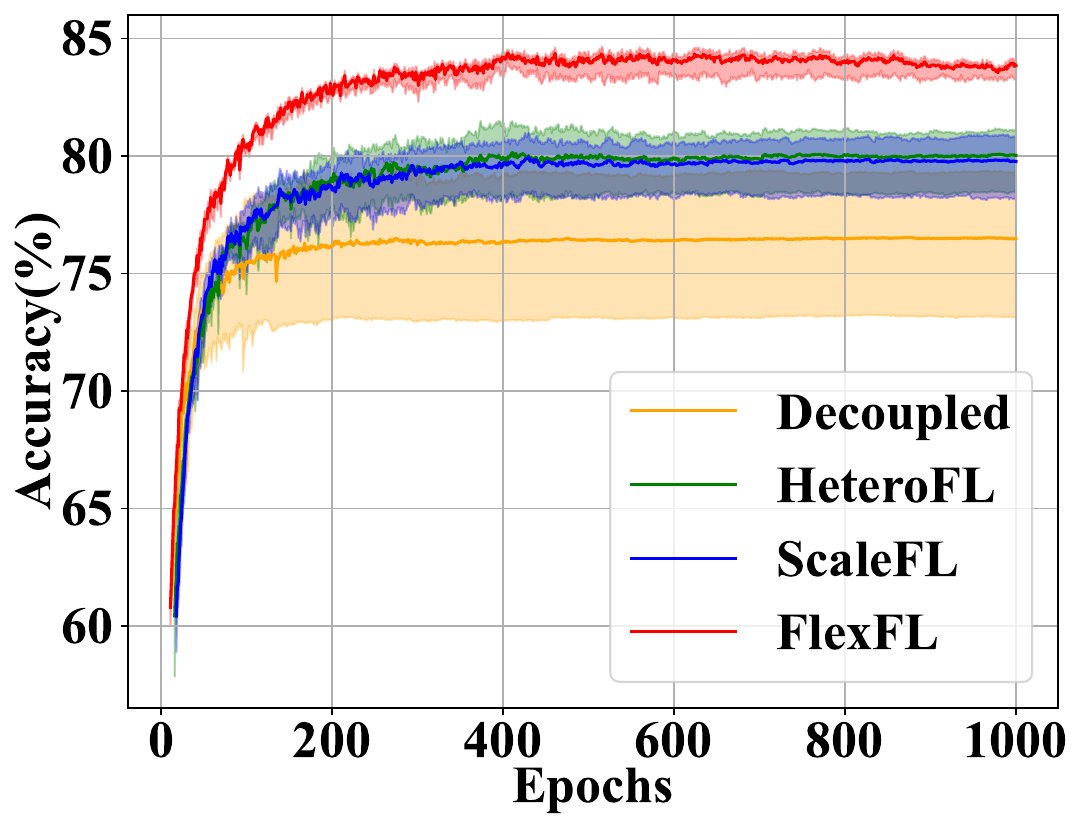}%
    }
    \hfil
    \subfloat[\footnotesize$f=10\%$]{\includegraphics[width=0.2\textwidth]{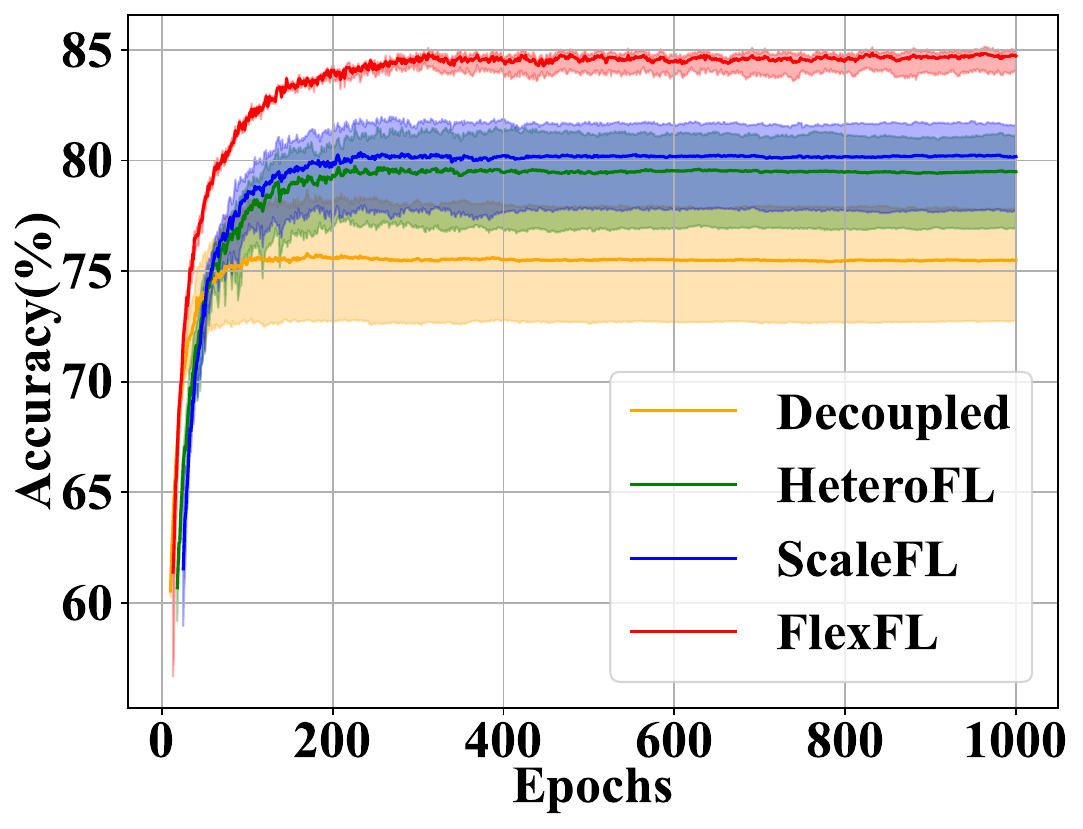}%
    }
     \vspace{-0.1in}
    \subfloat[\footnotesize$f=20\%$]{\includegraphics[width=0.2\textwidth]{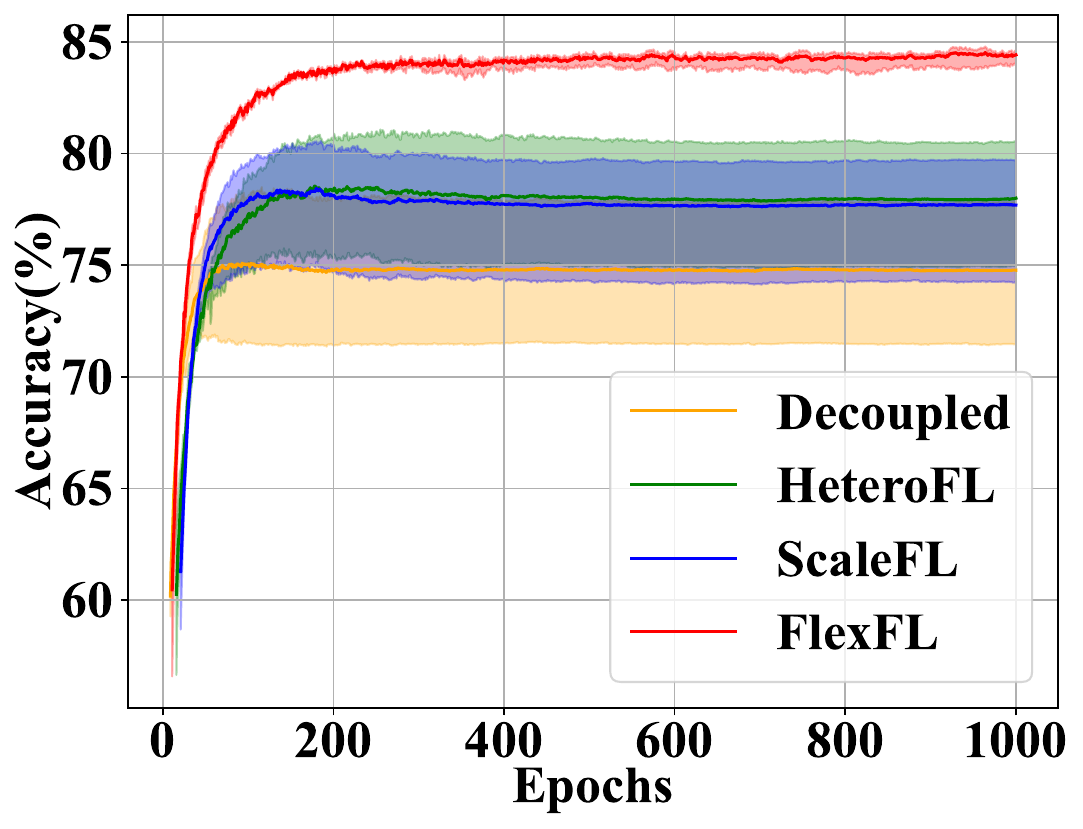}%
    }
    \hfil
    \subfloat[\footnotesize$f=50\%$]{\includegraphics[width=0.2\textwidth]{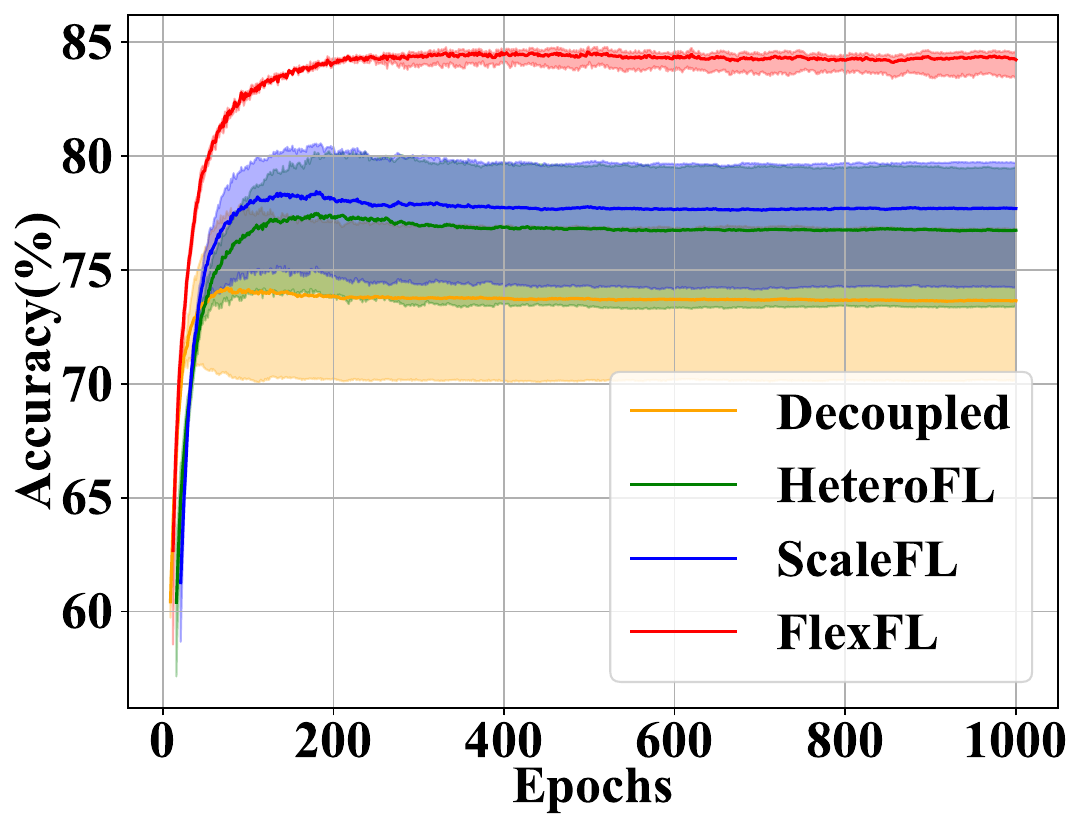}%
    }
\caption{Learning curves for different ratios of selected devices.}
\label{fig: different simultaneous devices}
\vspace{-0.20in}
\end{figure}

Assume that there are a total of $|C|=100$ devices. 
Figure \ref{fig: different simultaneous devices}
compares the training performance of  FL methods considering different 
ratios $f$ of selected devices based on CIFAR10 and VGG16 within an IID
scenario. 
We can find that our approach achieves the best performance in all four cases. Moreover, as the ratio $f$  increases, the accuracy of our method remains stable, and the differences between the accuracy of the large and small models are the smallest.
Conversely, for ScaleFL, as $f$ increases, its performance slightly decreases, and the difference in accuracy between the large and small models gradually increases.

\subsubsection{Proportions of Different Devices} 
\label{sec: Proportions}
\begin{figure}[h]
\vspace{-0.15in}
\centering
    \subfloat[\footnotesize$1:1:8$]{\includegraphics[width=0.2\textwidth]{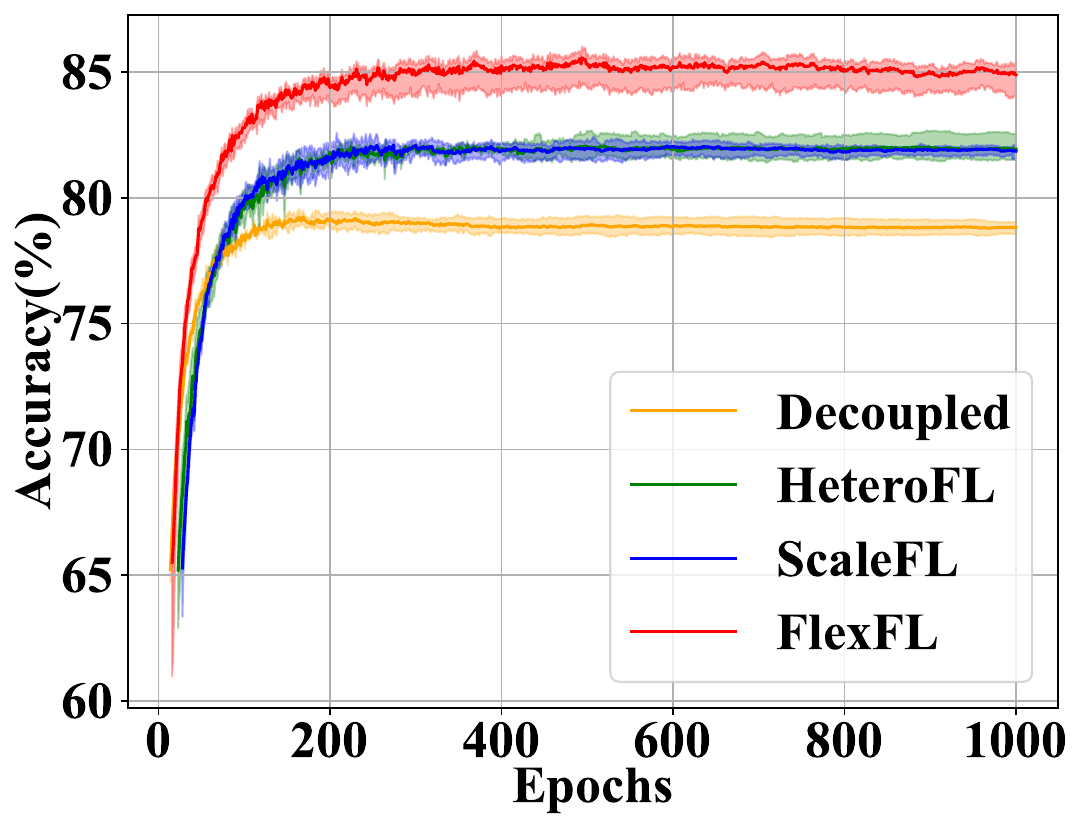}%
    }
    \hfil
    \subfloat[\footnotesize$1:8:1$]{\includegraphics[width=0.2\textwidth]{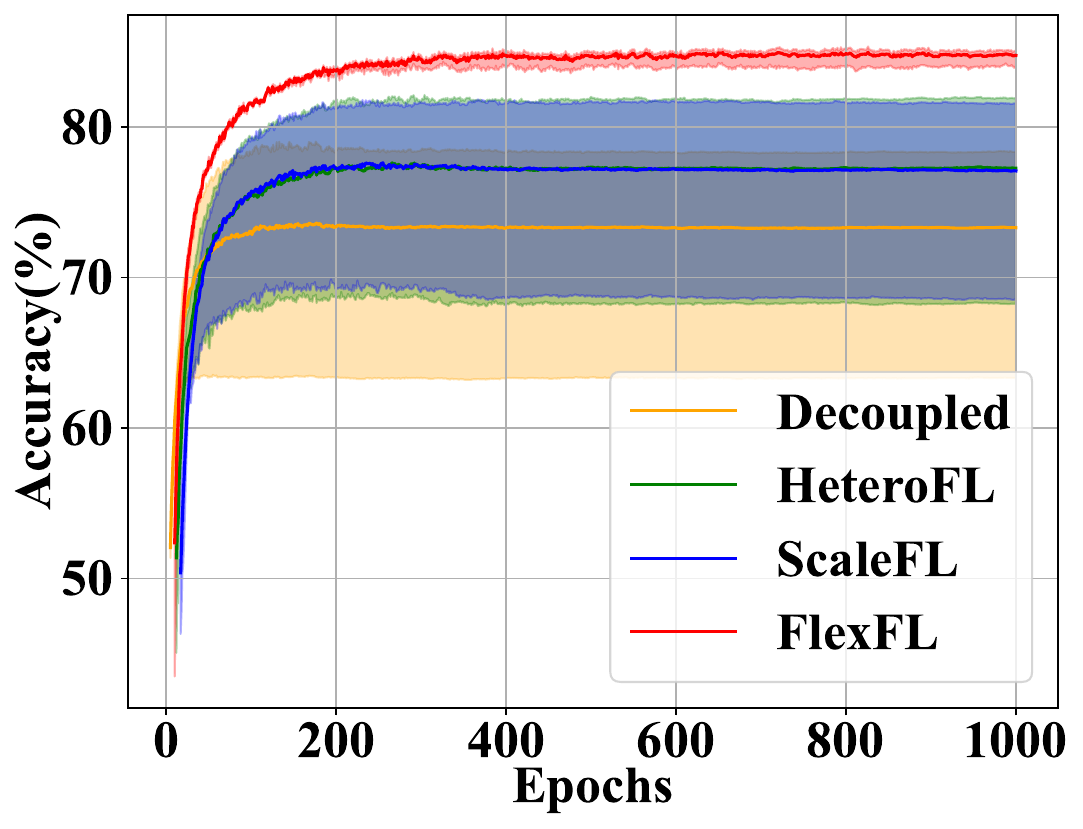}%
    }
     \vspace{-0.1in}
    \subfloat[\footnotesize$8:1:1$]{\includegraphics[width=0.2\textwidth]{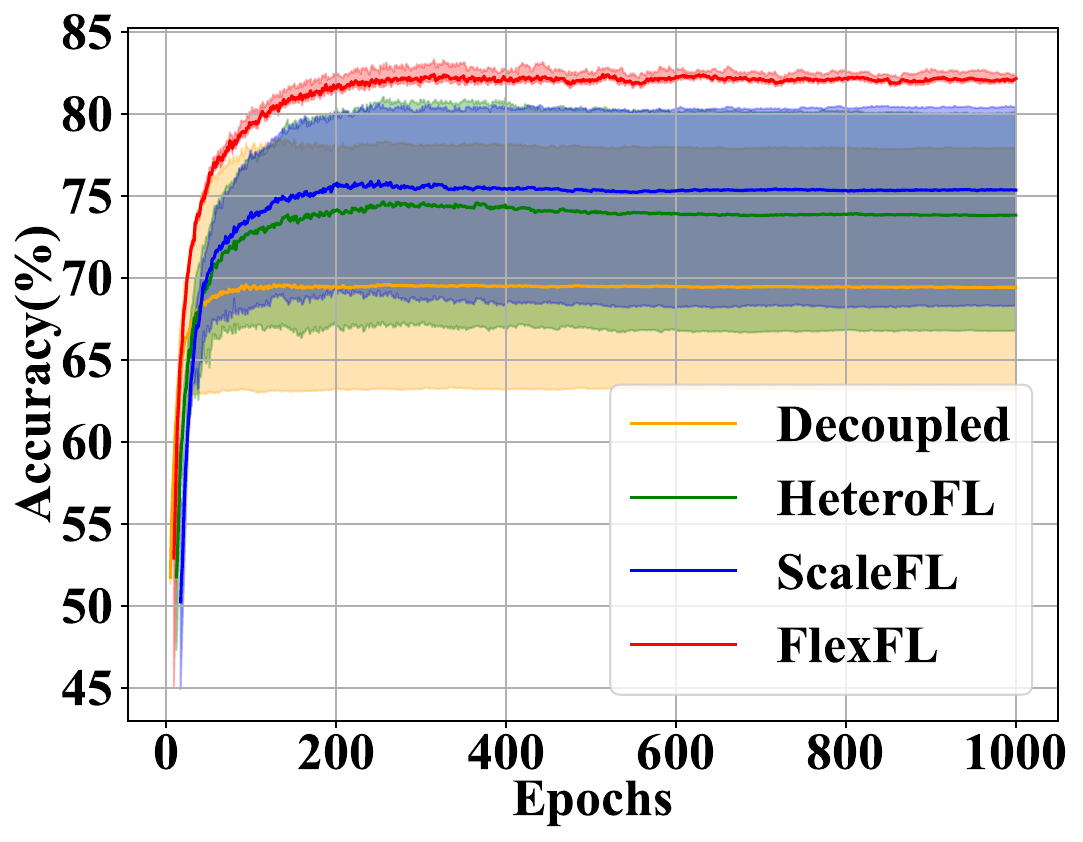}%
    }
    \hfil
    \subfloat[\footnotesize$4:3:3$]{\includegraphics[width=0.2\textwidth]{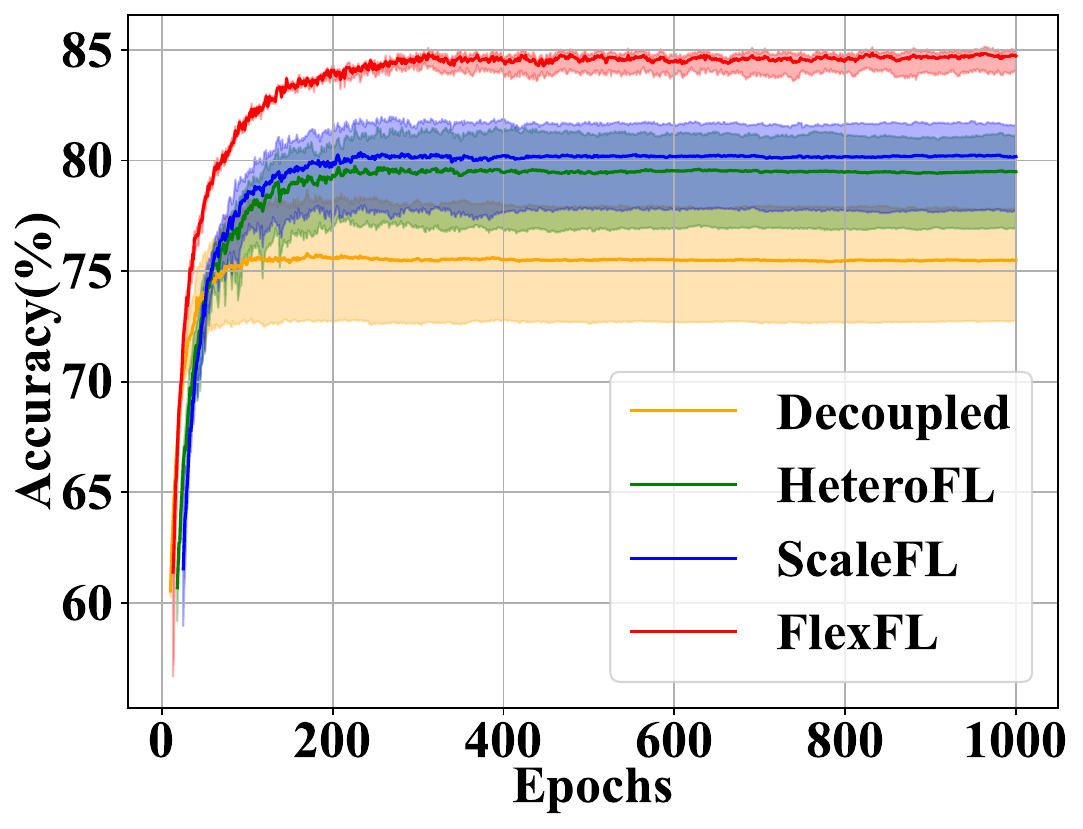}%
    }
\caption{Learning curves for different proportions (small:medium:large) of devices using VGG on CIFAR10 in the IID scenario.}
\label{fig: different proportion of devices}
\vspace{-0.15in}
\end{figure}

We compared our method with three baselines in terms of accuracy under different proportions of devices, as shown in Figure \ref{fig: different proportion of devices}.
We categorized all devices into three groups, and the uncertainty configurations for each group of devices are shown in Table~\ref{Table: uncertain devices}.
We varied the device proportions to 1:1:8, 1:8:1, and 8:1:1.
According to our experimental results, we observed that our method outperformed baselines in terms of average model accuracy across all device proportions. Furthermore, as the device proportions changed, our method's accuracy remained relatively stable, while other baselines exhibited performance degradation. Moreover, when the number of medium and small models was higher, the difference between the large and small model accuracy in our method was small. In contrast, the difference between the highest and lowest model accuracy in other Baselines reached approximately 10\%. 
This indicates that our method can effectively accommodate extreme device resource variations with different proportions.

\begin{table}[h]
\vspace{-0.15in}
\centering
\caption{Test accuracy comparison with different $\Gamma$.}
\vspace{-0.10in}
\footnotesize
\label{Table: adaptive size}
\begin{tabular}{ccc}
\hline
 $\Gamma$ & Global  Accuracy & Avg  Accuracy \\ \hline
1\%                         & 84.38\%             & 84.22\%           \\
2\%                        & 84.86\%            & 84.65\%          \\
5\%                        & 84.88\%            & 84.67\%          \\
10\%                          & 85.16\%               & 84.75\%             \\
15\%                        & 84.40\%              & 84.21\%            \\ \hline
\end{tabular}
\vspace{-0.15in}
\end{table}

\subsubsection{Adaptive Local Model Pruning Size} 
\label{sec: Adaptive Local Model Pruning Size}
We conducted experiments on CIFAR10 within an IID scenario with VGG16 to evaluate the impact of adaptive pruning sizes $\Gamma$. 
The experimental results are presented in Table \ref{Table: adaptive size}. We can find that the optimal value for the hyperparameter $\Gamma$ in our experiments is 10\%. 
%
%
This is mainly because a low value of $\Gamma$ leads to lower utilization of adaptive models, meaning more devices degrade their received models directly to smaller models. 
In contrast, although a high value of $\Gamma$ can improve the utilization of adaptive models, it causes smaller sizes of adaptive models, which results in a lower utilization of resources.
%

\subsubsection{Self-KD Hyperparameter Settings} 
\label{sec: Self-KD Hyperparameter Settings}
We investigated the impact of the hyperparameter $\lambda$ in self-KD on our experiments. 
In our experimental setup, we fixed the temperature parameter $\tau=3$ for self-KD and varied the coefficient of KL-loss $\lambda$ during local training as $\lambda=0,5,10,20,50$.

\begin{figure}[h]
\vspace{-0.1in}
\centering
\includegraphics[width=0.22\textwidth]{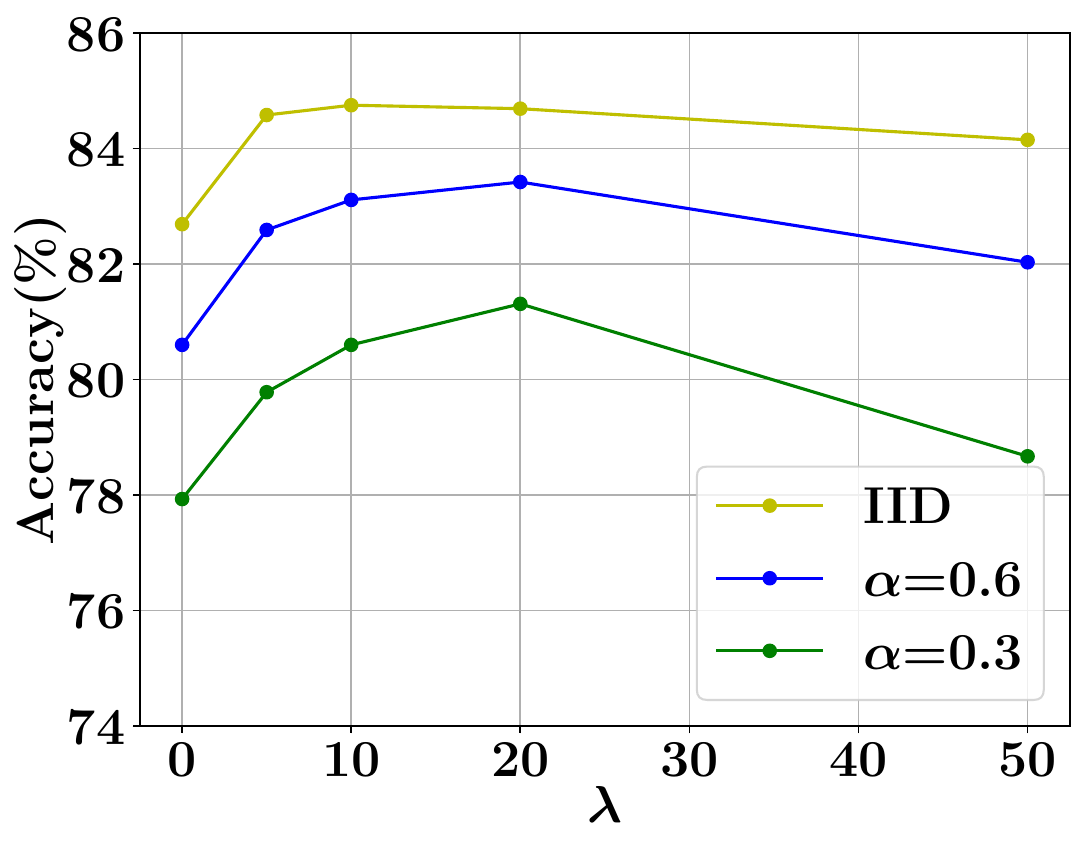}
\vspace{-0.1in}
\caption{Average accuracy  of VGG16 on CIFAR10 with different $\lambda$.}
\vspace{-0.1in}
\label{fig:  Distillation experiment}
\end{figure}

The experimental results in Figure \ref{fig: Distillation experiment} show that without self-KD ($\lambda=0$), the accuracy of our method decreased by approximately 2-4\% compared to when distillation was used. Furthermore, with increasing distillation coefficient $\lambda$, the accuracy showed an initial increase followed by a decreasing trend in both IID and Non-IID scenarios.
When the distillation coefficient $\lambda$ is at a reasonable range, i.e. $\lambda \in[10,20]$, the overall loss can be well balanced between distillation loss and cross-entropy loss. When $\lambda$ is set to a high value, the overall loss is dominated by distillation loss, which hinders effective learning of knowledge from the local dataset, resulting in an accuracy decrease.

\begin{figure*}[h]
\centering
    \subfloat[\footnotesize$\mathrm{IID}$]{\includegraphics[width=0.3\textwidth]{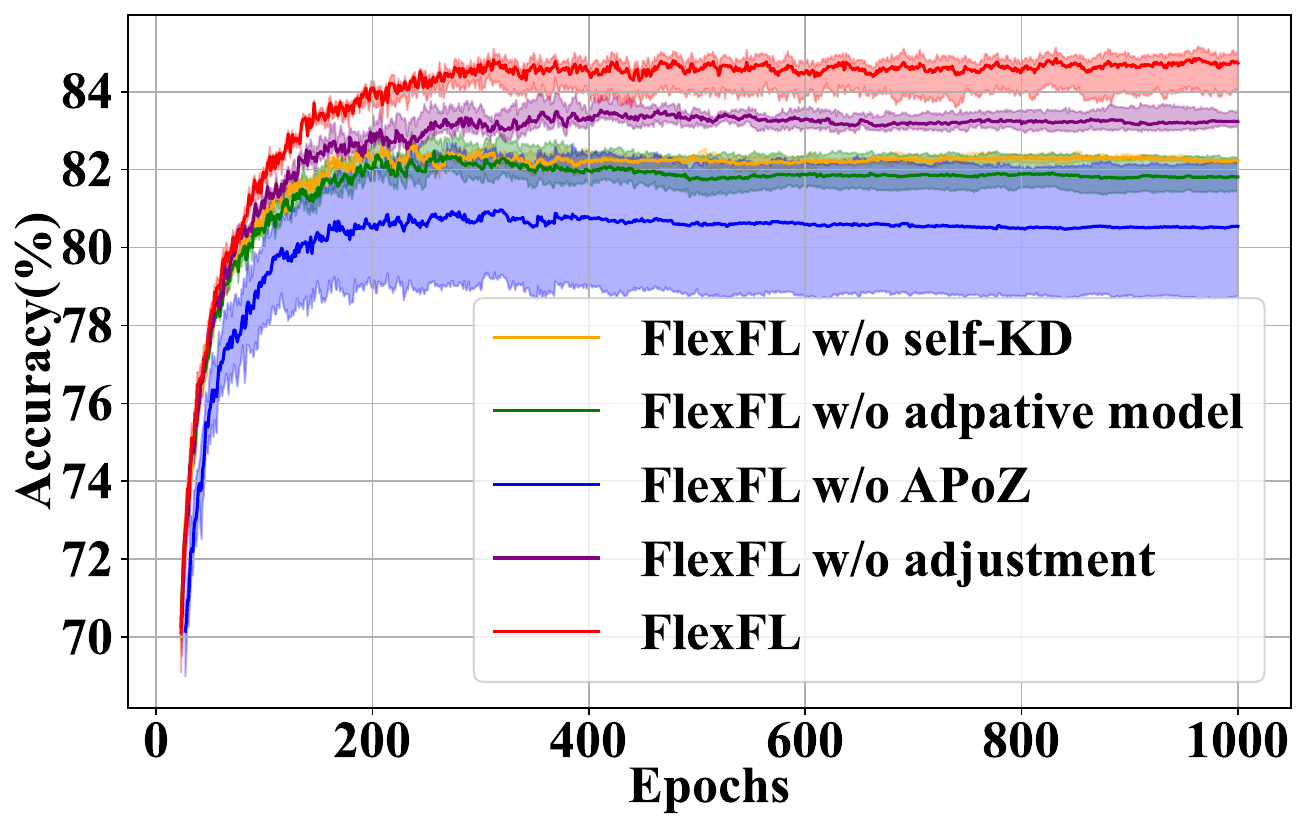}%
    }
    \hfil
    \subfloat[\footnotesize$\alpha = 0.6$]{\includegraphics[width=0.3\textwidth]{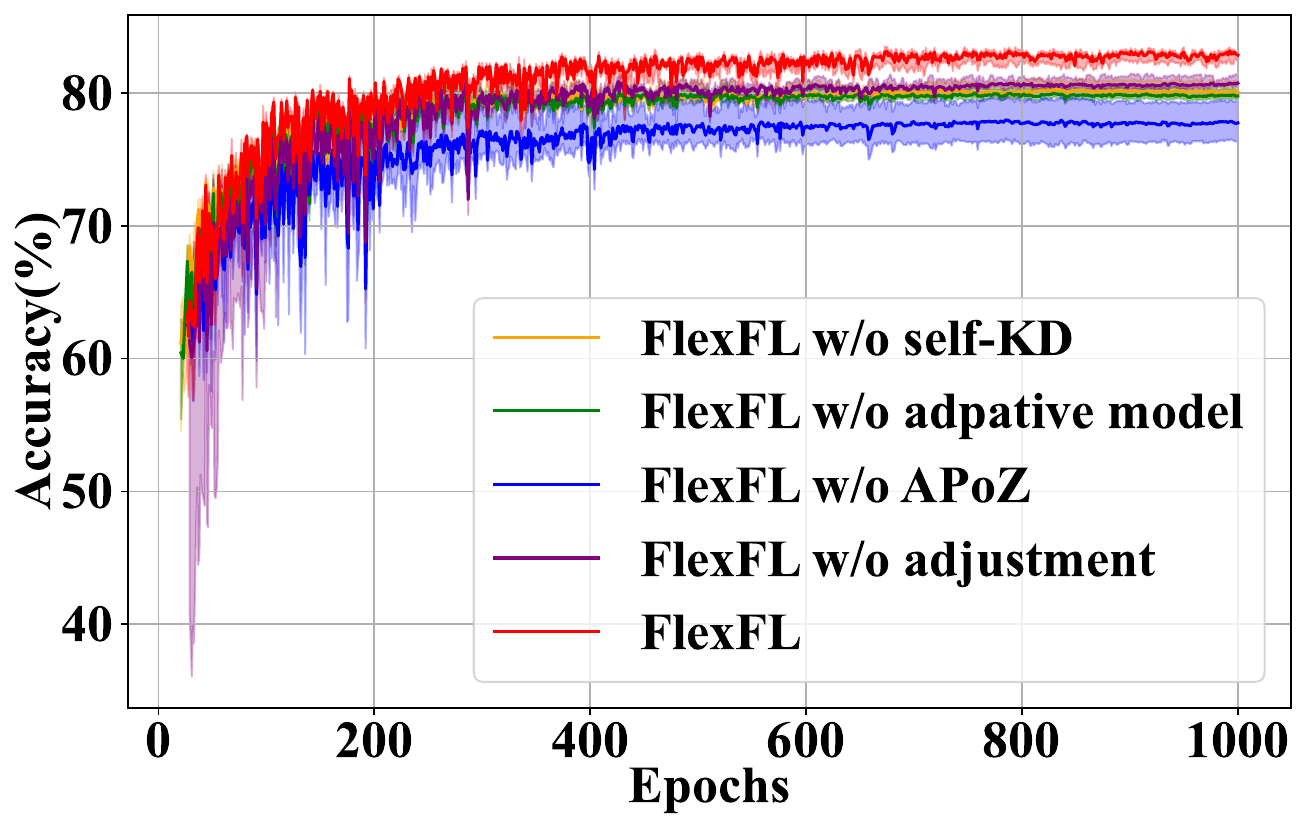}%
    }
    \hfil
    \subfloat[\footnotesize$\alpha = 0.3$]{\includegraphics[width=0.3\textwidth]{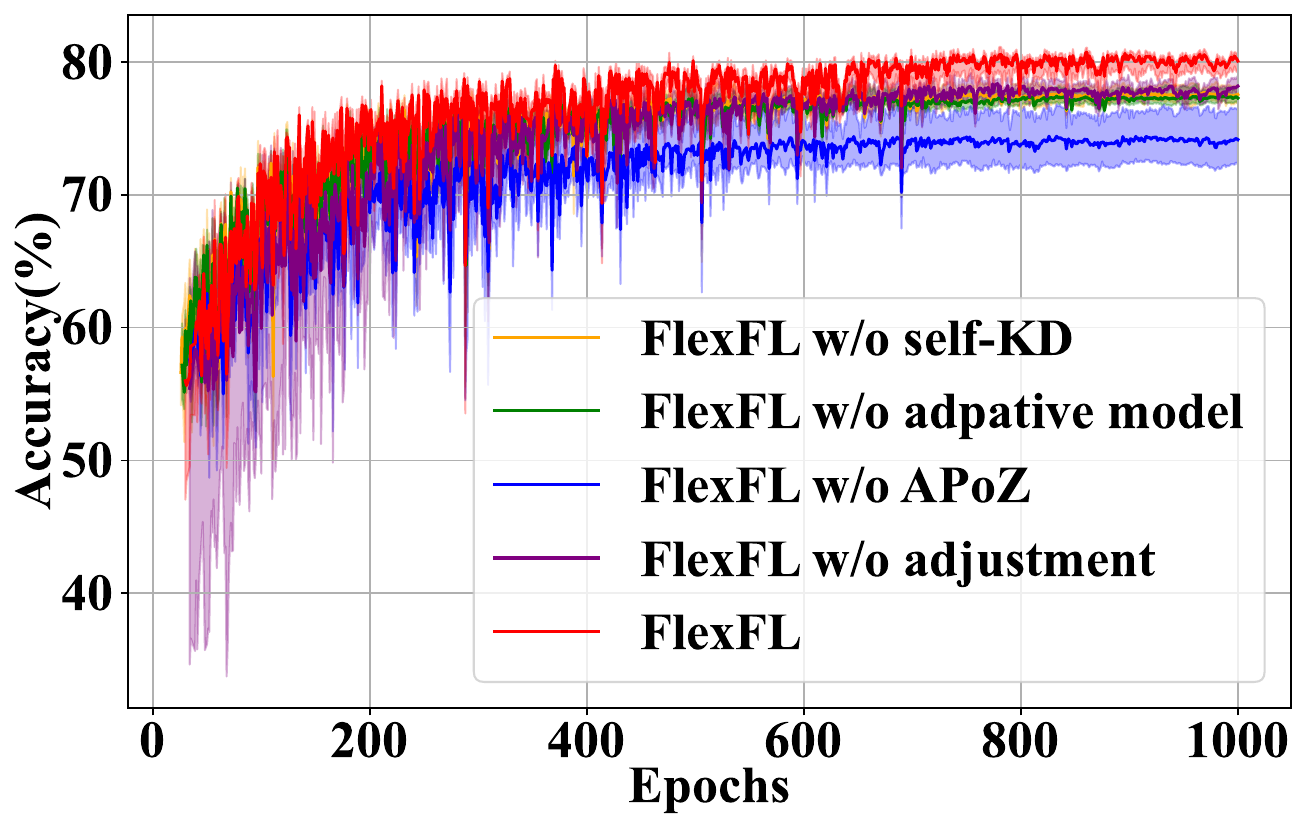}%
    }
\caption{Ablation study results for FlexFL (VGG on CIFAR10).}
\label{fig: ablation proportion of devices}
\vspace{-0.2in}
\end{figure*}

\begin{table}[h]
\vspace{-0.1in}
\centering
\caption{Configurations of different device resource distributions.}
\vspace{-0.05in}
\scriptsize
\label{Table: Resource distribution table}
\begin{tabular}{c|ccc}
\hline
 Configuration                      & \# Device & Max Capacity $r_M$ & Variance $u$      \\ \hline
\multirow{3}{*}{Conf1} & 40\%      & 35               & $\sigma^2 = 0$ \\
                       & 30\%      & 60               & $\sigma^2 = 0$\\
                       & 30\%      & 110              & $\sigma^2 = 0$ \\ \hline
\multirow{3}{*}{Conf2} & 40\%      & 35               & $\sigma^2 \in[5,8,10] $                     \\
                       & 30\%      & 60               & $\sigma^2 \in[5,8,10]  $                \\
                       & 30\%      & 110              & $\sigma^2 \in[5,8,10]  $                 \\ \hline
\multirow{3}{*}{Conf3} & 40\%      & 35               & $\sigma^2 \in[10,20,30]$ \\
                       & 30\%      & 60               & $\sigma^2 \in[10,20,30]$ \\
                       & 30\%      & 110              & $\sigma^2 \in[10,20,30]$ \\ \hline
\end{tabular}
\vspace{-0.05in}
\end{table}

\subsubsection{Different Settings of Resource Distributions} 
To explore our method's adaptability in different resource allocation scenarios, we constructed three distinct configuration plans, as shown in Table \ref{Table: Resource distribution table}.
\textit{Conf1} indicates a constant number of resources for each device, \textit{Conf2} indicates slight fluctuations in the resources of devices , and \textit{Conf3} suggests significant resource fluctuations across devices.

\begin{table}[h]
\vspace{-0.05in}
\centering
\caption{Test accuracy (\%) of models in different configurations.}
\vspace{-0.1in}
\scriptsize
\label{Table: Resource distribution experiment}
\begin{tabular}{c|c|ccc}
\hline
\multirow{2}{*}{Configuration} & \multirow{2}{*}{Algoroithm} & \multicolumn{3}{c}{Accuarcy}                                       \\ \cline{3-5} 
                              &                             & IID                  & $\alpha =0.6$        & $\alpha =0.3$        \\ \hline
  \multirow{4}{*}{Conf1}        & Decoupled                   & 75.83/72.00          & 73.20/69.62          & 70.23/65.20          \\
                              & HeteroFL                    & 79.08/76.21          & 76.91/74.27          & 73.98/69.96          \\
                              & ScaleFL                     & 80.14/77.52          & 76.71/74.15          & 72.24/69.27          \\
                              & FlexFL                      & \textbf{84.46/84.64} & \textbf{83.36/83.70} & \textbf{80.70/81.17} \\ \hline
\multirow{4}{*}{Conf2}        & Decoupled                   & 75.81/72.76          & 73.57/69.91          & 69.95/66.01          \\
                              & HeteroFL                    & 79.75/76.92          & 77.53/75.45          & 73.89/71.50          \\
                              & ScaleFL                     & 80.36/77.72          & 75.99/74.00          & 72.95/69.87          \\
                              & FlexFL                      & \textbf{84.75/85.16} & \textbf{83.11/83.45} & \textbf{80.60/81.06} \\ \hline

\multirow{4}{*}{Conf3}        & Decoupled                   & 76.90/73.16          & 74.38/69.42         & 71.35/66.27          \\
                              & HeteroFL                    & 79.23/77.67          & 76.49/73.16          & 73.12/71.36          \\
                              & ScaleFL                     & 80.24/78.79          & 76.37/74.96          & 73.08/71.50          \\
                              & FlexFL                      & \textbf{84.31/84.74} & \textbf{83.70/84.03} & \textbf{81.21/81.58}          \\ \hline
\end{tabular}
\vspace{-0.15in}
\end{table}

We conducted a study comparing the accuracy differences between our method and three baseline methods, with results shown in 
Table \ref{Table: Resource distribution experiment}.
Our method achieved approximately a 4\% performance improvement compared to ScaleFL across all configs, indicating that our method can maintain high accuracy when dealing with various degrees of resource fluctuations.

\subsubsection{Real-world Datasets}
To validate the generalization ability of our approach, we extended our experiments to include real-world datasets, i.e., FEMNIST~\cite{LEAF} and Widar~\cite{fedaiot}, in addition to image recognition datasets.
The FEMNIST dataset comprises 180 devices, with each training round selecting 10\% devices. The data distribution on devices is naturally non-IID.
We assumed that the Widar dataset involves 100 devices following given Dirichlet distributions, and 10 devices are selected for local training in each FL round. 
We applied the uncertainty settings in Table \ref{Table: uncertain devices}  to all devices.

\begin{table}[h]
\vspace{-0.1in}
\centering
\caption{Test accuracy (\%) comparison on Real-world datasets.
}
\vspace{-0.1in}
\scriptsize
\label{Table: real-world dataset}
\addtolength{\tabcolsep}{-2pt}
\begin{tabular}{cc|c|ccc}
\hline
\multirow{2}{*}{Model}       & \multirow{2}{*}{Algorithm} & \multirow{2}{*}{FEMNIST}               & \multicolumn{3}{c}{WIDAR}                                          \\ \cline{4-6} 
                             &                            &                     & IID                  & $\alpha = 0.6$       & $\alpha = 0.3$       \\ \hline
\multirow{4}{*}{VGG16}       & Decoupled                  & 78.70/71.63          & 67.51/60.80          & 66.41/60.28          & 64.12/58.47          \\
                             & HeteroFL                   & \textbf{79.84}/72.54 & 70.81/67.35          & 68.82/64.90          & \textbf{67.28}/64.11 \\
                             & ScaleFL                    & 70.85/63.61          & 69.99/67.80          & 68.42/65.95          & 64.50/62.99          \\
                             & FlexFL                     & 77.16/\textbf{75.94}          & \textbf{71.66/72.13} & \textbf{71.04/70.92} & 67.20/\textbf{68.11}          \\ \hline
\multirow{4}{*}{Resnet34}    & Decoupled                  & 74.02/64.27          & 63.54/58.36          & 60.65/54.89          & 58.84/53.84          \\
                             & HeteroFL                   & 76.99/68.07          & 67.90/63.05          & 65.17/60.59          & 59.83/56.71          \\
                             & ScaleFL                    & 76.94/69.25          & 64.77/61.17          & 61.44/58.92          & 58.97/57.67          \\
                             & FlexFL                     & \textbf{83.64/80.46} & \textbf{72.43/73.10} & \textbf{71.19/71.57} & \textbf{67.91/68.77} \\ \hline
\multirow{4}{*}{MobileNetV2} & Decoupled                  & 68.40/56.33          & 51.28/44.45          & 45.92/44.04          & 43.76/38.92          \\
                             & HeteroFL                   & 70.94/61.87          & 56.49/53.05          & 51.74/49.15          & 47.67/45.02          \\
                             & ScaleFL                    & 71.38/62.02          & 58.17/55.01          & 53.13/47.45          & 46.93/44.12          \\
                             & FlexFL                     & \textbf{77.25/71.38} & \textbf{64.92/65.67} & \textbf{63.43/63.98} & \textbf{57.61/59.26} \\ \hline
\end{tabular}
\vspace{-0.15in}
\end{table}

The results presented in Table \ref{Table: real-world dataset} demonstrate the performance of our method on ResNet34 and MobileNetV2, with performance improvements of up to 10.63\%.
There is minimal difference between the average model accuracy and the accuracy of the best-performing model. 
On VGG16, although FlexFL exhibits a 2.68\% lower average accuracy compared to HeteroFL on the FEMNIST dataset, FlexFL still demonstrates improved accuracy for the largest model.

\subsection{Ablation Study}
We conducted a study on the effectiveness of each component within our method to investigate their respective impacts on the accuracy of our approach.
We designed four varieties of FlexFL: 
i)
``w/o self-KD'' indicates the absence of self-distillation during local training;  ii)
``w/o adaptive model'' implies the utilization of only models $M_1$, $M_2$, and $M_3$, with the current model $M_i$ being pruned to the model $M_{i-1}$ under resource constraints;
iii)
``w/o APoZ'' involves only using adjustment weight $AdjW$ for model pruning; and iv)
``w/o adjustment'' entails utilizing APoZ for model pruning without adjustment weight $AdjW$.
Figure \ref{fig: ablation proportion of devices} shows 
the superiority of FlexFL against its four variants, 
indicating that the absence of our proposed components will decrease model accuracy, with the lack of APoZ causing the most significant decline.

\subsection{Evaluation on Real Test-bed}

To demonstrate the effectiveness of FlexFL in real AIoT scenarios, we conducted experiments on a real test-bed platform, which consists of 17 different AIoT devices and a cloud server.
Table~\ref{table: Composition of real test-bed platform} shows the details of the configuration of these AIoT devices.
Based on our real test-bed platform, we conducted experiments with a non-IID $\alpha =0.1$ scenario on CIFAR10~\cite{CIFAR} dataset using MobileNetV2~\cite{mobilenetv2} models and selected 10 devices to participate in local training in each FL training round.
We set a uniform time limit of 70000 seconds for FlexFL, ScaleFL, and HeteroFL.

\begin{table}[h]
\vspace{-0.1in}
\centering
\caption{Real test-bed platform configuration.}
\vspace{-0.1in}
\scriptsize
\label{table: Composition of real test-bed platform}
\begin{tabular}{c|cl|c|c}
\hline
 \textbf{Device}   & \multicolumn{2}{c|}{\textbf{Comp}}        & \textbf{Mem} & \textbf{Num} \\ \hline
 Raspberry Pi 4B   & \multicolumn{2}{c|}{ARM Cortex-A72 CPU}   & 2G           & 4            \\
 Jetson Nano       & \multicolumn{2}{c|}{128-core Maxwell GPU} & 8G           & 10           \\
 Jetson Xavier AGX & \multicolumn{2}{c|}{512-core NVIDIA GPU}  & 32G          & 3            \\ \hline
 Workstation       & \multicolumn{2}{c|}{NVIDIA RTX 4090 GPU}  & 64G          & 1            \\ \hline
\end{tabular}
\vspace{-0.15in}
\end{table}

\begin{figure}[h]
\vspace{-0.25 in}
\centering
    \subfloat[\footnotesize{Real Test-bed Platform}]{\includegraphics[width=0.22\textwidth]{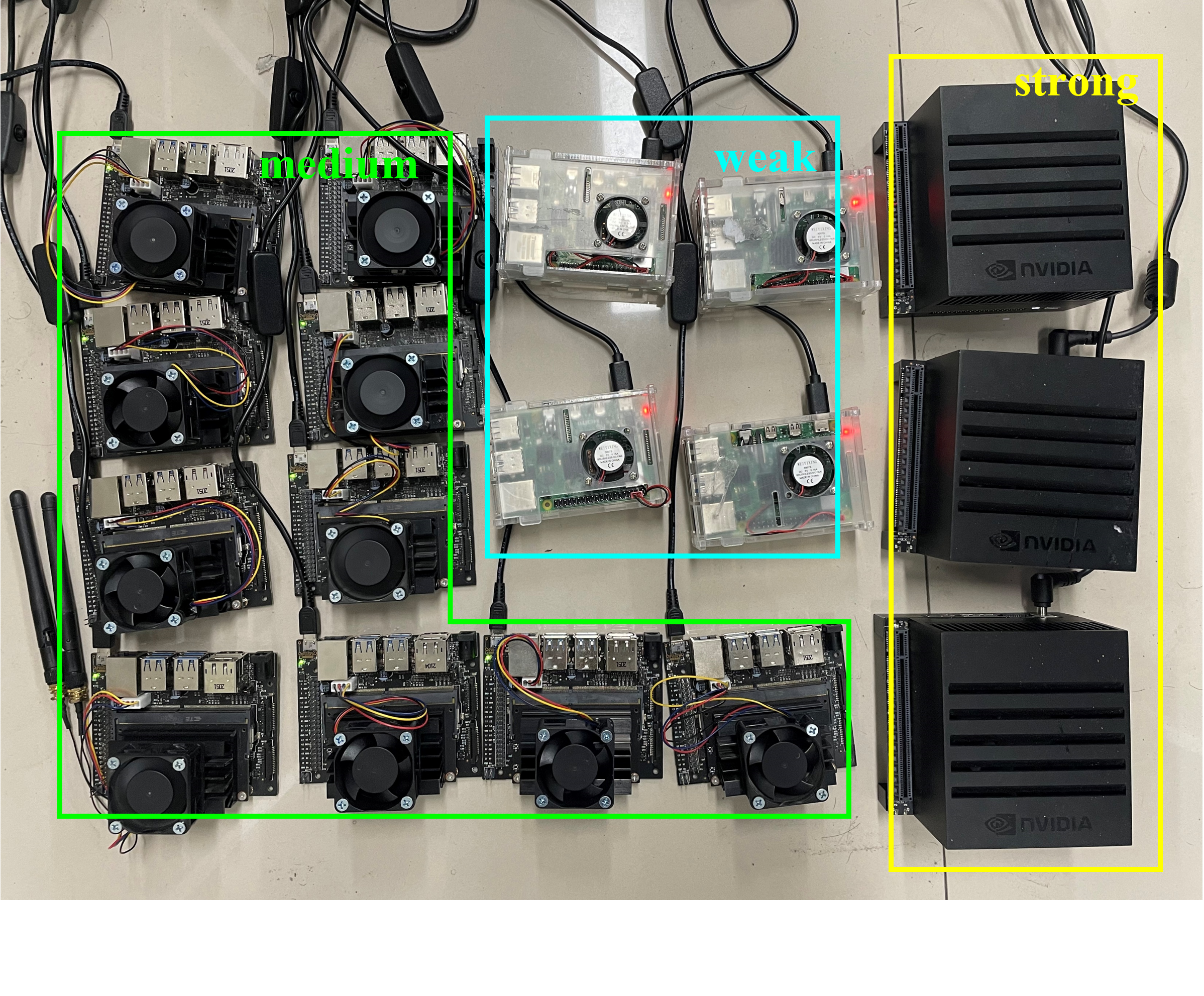}%
    }
    \hfil
    \subfloat[\footnotesize Learning Curves]{\includegraphics[width=0.24\textwidth]{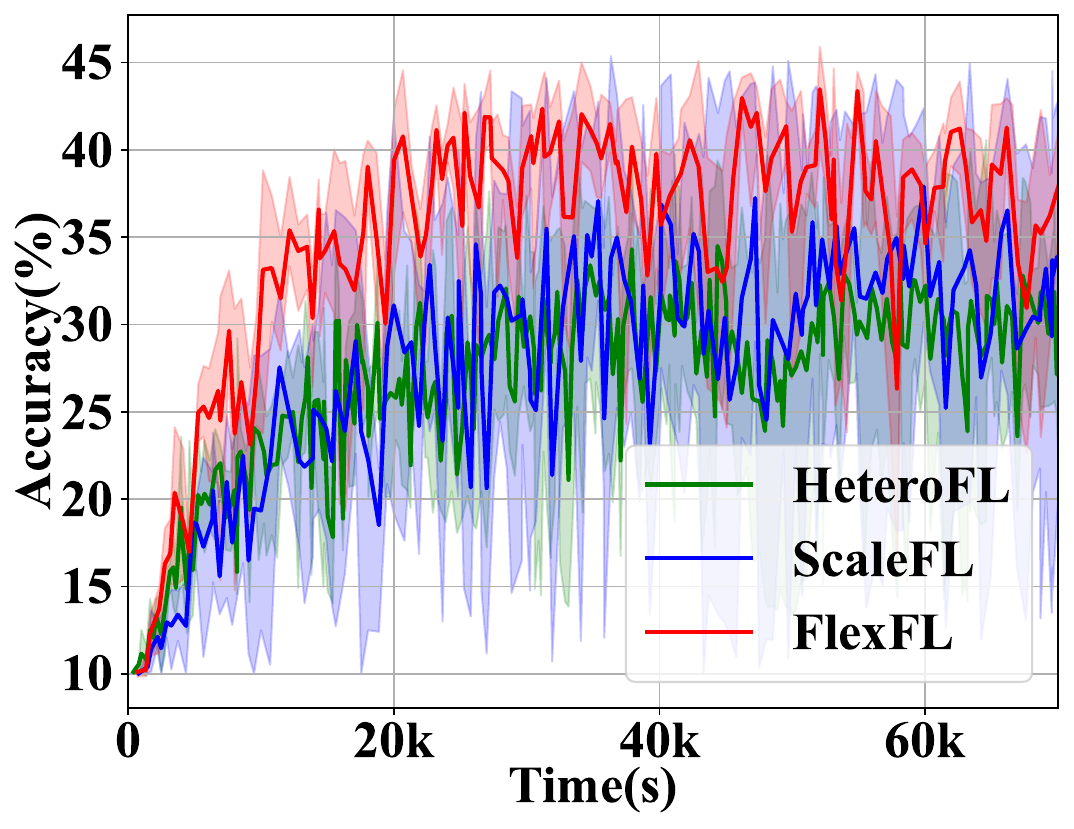}%
    }
\caption{Real test-bed devices and the learning curves.}
\label{fig: real}
\vspace{-0.25in}
\end{figure}

Figure \ref{fig: real} illustrates our real test-bed devices and the learning curves of the methods. 
From Figure~\ref{fig: real} (b), we can observe that FlexFL achieves the highest accuracy compared to ScaleFL and HeteroFL.
In addition, we can also find that compared with the two baselines, FlexFL has relatively small accuracy fluctuations.
Therefore, compared to ScaleFL and HereroFL, FlexFL still achieves the best inference accuracy and stability on the real test-bed platform.

\subsection{Discussion}
\subsubsection{Scalability Analysis}
\label{sec: Scalability Analysis}
With the increasing number of heterogeneous devices integrated into complex IoT systems, scalability becomes crucial for the deployment of FlexFL.
We conducted experiments to evaluate the scalability of FlexFL in Sections \ref{sec: Participating}, \ref{sec: Simultaneously}, and \ref{sec: Proportions}. 
We verified the performance of FlexFL using 500 devices for heterogeneous federated learning, selecting 50\% devices for training in each federated learning round, and considering scenarios with extreme variations in device resource distribution. The experimental results show that our approach can accommodate large FL applications with various network architectures. 

\begin{table}[h]
\centering

\vspace{-0.05in}
\caption{Time overhead of components per round (VGG ON IID CIFAR10).}
\label{Table: FlexFL distillation and adaptive pruning overhead ablation}
\scriptsize
\begin{tabular}{c|c|ccc}
\hline
\multirow{2}{*}{Method}                                                                            & \multirow{2}{*}{\begin{tabular}[c]{@{}c@{}}Dispatched  \\ Model Size\end{tabular}} & \multirow{2}{*}{\begin{tabular}[c]{@{}c@{}}Adaptive  \\ Pruning \end{tabular}}  & \multirow{2}{*}{\begin{tabular}[c]{@{}c@{}}Local  \\ Training\end{tabular}}      & \multirow{2}{*}{Total}  \\
& & & & \\ \hline
\multirow{4}{*}{FlexFL}                                                           & 100\%          & 1.45s (1.3\%)  & 108.59s (98.7\%) & 110.04s \\
                                                                                  & 50\%           & 1.45s (1.9\%)  & 73.92s (98.1\%)  & 75.37s  \\
                                                                                  & 25\%           & 0s (0\%)       & 46.06s (100\%)   & 46.06s  \\ \cline{2-5} 
                                                                                  & Avg       & 1.45s (2.0\%) & 70.37s (98.0\%)  & 71.82s  \\ \hline
\multirow{4}{*}{\begin{tabular}[c]{@{}c@{}}FlexFL  \\ w/o self-KD\end{tabular}} & 100\%          & 1.44s (2.2\%) & 64.17s (97.8\%)  & 65.61s  \\
                                                                                  & 50\%           & 1.45s (2.6\%) & 52.88s (97.4\%)  & 54.33s  \\
                                                                                  & 25\%           & 0s (0\%)             & 45.77s (100\%)   & 45.77s  \\ \cline{2-5} 
                                                                                  & Avg        & 1.44s (2.8\%) & 50.61s (97.2\%)  & 52.05s  \\ \hline
\end{tabular}
\end{table}

\subsubsection{Computation Overhead} 
\label{sec : Training Time and Communication Overhead Analysis}
To evaluate the computation overhead of components (i.e., pre-processing, adaptive local pruning, and self-KD local training) introduced by FlexFL, we conducted various experiments in our real test-bed platform to investigate their impacts on the overall training time. 
Specifically, we evaluated from two aspects: i) time overhead of components per FL training round; and ii) training time to achieve a specific test accuracy. 
From Table~\ref{Table: FlexFL distillation and adaptive pruning overhead ablation}, we can find that ``FlexFL without self-KD'' needs 52.05s on average, while FlexFL needs 71.82s on average. The introduction of self-KD will result in longer local training time. 
Note that our method's pre-processing time accounts for approximately 0.3\% of the total training time (219s for 1000 rounds of training), and its adaptive pruning time accounts for about 2\% of the local time.
Overall, the computational overhead of adaptive pruning and pre-training is almost negligible, and the main additional computational overhead of FlexFL comes from self-KD.
However, as shown in Table \ref{Table: overhead},
FlexFL needs much less training time to achieve a specific test accuracy than HeteroFL and ScaleFL. Specifically, compared with ScaleFL, FlexFL can achieve a training speedup of up to 6.63$\times$.
Here, ``All Large'' represents the results obtained by only dispatching the largest models for FL training under FedAvg, and the notation ``N/A'' means not available.
Moreover, FlexFL can achieve an accuracy of 85\%, while all the other methods fail, mainly benefit from the performance improvement brought about by our proposed self-KD technique.

\begin{table}[h]
\centering

\caption{Training Time and communication overhead to achieve the same accuracy (VGG ON IID CIFAR10).}
\scriptsize
\label{Table: overhead}
\begin{tabular}{cc|cccc}
\hline
\multicolumn{2}{c|}{Target Accuracy}                                                                     & 70\%    & 75\%    & 80\%     & 85\%     \\ \hline
\multicolumn{1}{c|}{\multirow{3}{*}{FlexFL}}                                                            & Time (s)      & 4849  & 6857  & 11736  & 32810  \\
\multicolumn{1}{c|}{}                                                                                   & Dispatch (MB) & 18337 & 26580 & 45860  & 132298 \\
\multicolumn{1}{c|}{}                                                                                   & Upload (MB)   & 17751 & 25914 & 44736  & 129075 \\ \hline
\multicolumn{1}{c|}{\multirow{3}{*}{\shortstack{FlexFL\\ w/o self-KD}}} & Time (s)      & 2566  & 4013  & 7985   & N/A      \\
\multicolumn{1}{c|}{}                                                                                   & Dispatch (MB) & 16520 & 28767 & 59857  & N/A      \\
\multicolumn{1}{c|}{}                                                                                   & Upload (MB)   & 16022 & 27973 & 58208  & N/A      \\ \hline
\multicolumn{1}{c|}{\multirow{3}{*}{ScaleFL}}                                                           & Time (s)      & 9566  & 14808 & 77926  & N/A      \\
\multicolumn{1}{c|}{}                                                                                   & Dispatch (MB) & 35935 & 55727 & 300499 & N/A      \\
\multicolumn{1}{c|}{}                                                                                   & Upload (MB)   & 34252 & 53226 & 288020 & N/A      \\ \hline
\multicolumn{1}{c|}{\multirow{3}{*}{HeteroFL}}                                                          & Time (s)      & 4900  & 8328  & N/A      & N/A     \\
\multicolumn{1}{c|}{}                                                                                   & Dispatch (MB) & 26412 & 45624 & N/A      & N/A      \\
\multicolumn{1}{c|}{}                                                                                   & Upload (MB)   & 24023 & 41822 & N/A      & N/A      \\ \hline
\multicolumn{1}{c|}{\multirow{3}{*}{All Large}}                                                          & Time (s)     & 3821  & N/A     & N/A      & N/A      \\
\multicolumn{1}{c|}{}                                                                                   & Dispatch (MB) & 37684 & N/A     & N/A      & N/A      \\
\multicolumn{1}{c|}{}                                                                                   & Upload (MB)   & 37684 & N/A     & N/A      & N/A      \\ \hline
\end{tabular}
\end{table}

\subsubsection{Communication Overhead}
\label{sec: overhead ablation}
From Table~\ref{Table: overhead}, we can observe that FlexFL achieves the optimal communication overhead under all target accuracy levels and can reduce communication overhead (Dispatch+Upload) by up to 85\%. Since APoZ-guided pruning, adaptive local pruning, and self-distillation do not rely on any extra complex data structures, the additional memory overhead they introduce is negligible. 
Note that APoZ-guided pruning is performed on the server side and only once upon initialization, it does not impose any burden on devices.

%% file: conclusion.tex
\section{Conclusion}\label{section: conclusion}

Due to the lack of strategies 
to generate high-performance heterogeneous models, 
existing heterogeneous FL suffers from low inference performance, 
especially for various uncertain scenarios.
To address this problem, this paper presents a novel heterogeneous FL approach named FlexFL, which adopts an APoZ-guided flexible pruning strategy to wisely generate heterogeneous models to fit various heterogeneous AIoT devices.
Based on our proposed  adaptive local pruning mechanism, 
FlexFL enables devices to further prune their received 
models to accommodate various uncertain scenarios. Meanwhile, 
FlexFL introduces an effective self-knowledge distillation-based local training strategy, which can improve the inference capability of large models by learning from small models, thus boosting the overall FL performance.  
Comprehensive experimental results
obtained from simulation- and real test-bed-based AIoT systems show that our approach can achieve better inference performance
compared with state-of-the-art heterogeneous FL methods.



%% file: ack.tex
\section{Acknowledgement}
This work was supported by Natural Science Foundation of China (62272170),   and ``Digital Silk Road'' Shanghai International Joint Lab of Trustworthy Intelligent Software (22510750100), Shanghai Trusted Industry Internet Software Collaborative Innovation Center, the National Research Foundation, Singapore, and the Cyber Security Agency under its National Cybersecurity R\&D Programme (NCRP25-P04-TAICeN).
Any opinions, findings and conclusions or recommendations expressed in this material are those of the author(s) and do not reflect the views of National Research Foundation, Singapore and Cyber Security Agency of Singapore.